\documentclass[11pt,english]{article}

\usepackage[utf8]{inputenc} 

\usepackage{dsfont,mathrsfs}

\usepackage[left=1in,right=1in,top=1in,bottom=1in]{geometry}
\usepackage{authblk}
\usepackage{enumitem}
\usepackage{ulem} 
\usepackage{empheq}

\usepackage{graphicx}
\usepackage{subcaption}
\usepackage{float}
\usepackage{wrapfig} 
\usepackage{diagbox}

\usepackage{xcolor}
\usepackage{color}
\definecolor{darkgreen}{rgb}{0,0.5,0}
\definecolor{darkblue}{rgb}{0,0,0.6}
\definecolor{purple}{rgb}{0.4,.2,0.7}

\usepackage{amsmath}
\usepackage{amssymb}

\usepackage{tensor} 
\usepackage{slashed}
\usepackage{mathtools}
\usepackage{leftidx}
\usepackage{esint}
\usepackage{amsfonts,amsthm,bm}

\usepackage{physics}
\usepackage{feynmf}
\usepackage{braket}

\usepackage{prettyref}
\usepackage[colorlinks=true,citecolor=darkgreen,linkcolor=purple,urlcolor=purple]{hyperref}

\usepackage[numbers,sort&compress]{natbib}

\usepackage{tikz}

\newcommand{\nn}{\nonumber}

\def\p{\partial}
\def\half{\frac{1}{2}}
\def\be{\begin{equation}}
\def\ee{\end{equation}}
\def\ba{\begin{aligned}}
\def\ea{\end{aligned}}
\def\mc{\mathcal}
\def\eps{\varepsilon}

\def\lam{\lambda}

\def\Gam{\Gamma}
\def\gam{\gamma}

\def\integ{\mathbb{Z}}
\newcommand{\inner}[2]{\langle \, #1 \mid #2 \, \rangle}

\numberwithin{equation}{section}
\numberwithin{figure}{section}
\numberwithin{table}{section}

\begin{document}

\title{\LARGE\textsc{Dynamical Edge Modes in $p$-form Gauge Theories}}


\author[a]{\vskip1cm \normalsize Adam Ball}
\affil[a]{\it \normalsize Perimeter Institute for Theoretical Physics, Waterloo, ON, Canada}
\author[b]{\normalsize Y.T.\ Albert Law}
\affil[b]{\it \normalsize Stanford Institute for Theoretical Physics, Stanford, CA, USA}

\date{}
\maketitle

\begin{center}
	\vskip-10mm
	{\footnotesize \href{mailto:aball1@pitp.ca}{aball1@pitp.ca}\,,\;  \href{mailto:ytalaw@stanford.edu}{ytalaw@stanford.edu}} 
\end{center}

\vskip10mm

\thispagestyle{empty}

\begin{abstract}

We extend our recently identified dynamical edge mode boundary condition to $p$-form gauge theories, revealing their edge modes as Goldstone bosons arising from gauge transformations with support on the boundary. The symplectic conjugates of these edge modes correspond to the electric-field-like components normal to the boundary. We demonstrate that both the symplectic form and the Hamiltonian naturally decompose into bulk and edge parts. When the boundary is a stretched horizon, we show that the thermal edge partition function reduces to that of a codimension-two ghost $(p-1)$-form residing on the bifurcation surface. These findings provide a dynamical framework that elucidates observations made by several authors. Additionally, we generalize Donnelly and Wall's non-dynamical approach to obtain edge partition functions for both massive and massless $p$-forms. In the context of a de Sitter static patch, these results are consistent with the edge partition functions found by several authors in arbitrary dimensions.

\end{abstract}

\newpage

\tableofcontents


\section{Introduction}

Degrees of freedom associated with boundaries are ubiquitous. In gauge theory and gravity, ``edge modes'' have been recognized as the underlying mechanism responsible for several universal phenomena that, although distinct and seemingly unrelated, share a common origin.\footnote{Some aspects of this phenomenon extend to {\it any} theories with spin $s \geq 1$, both massive and massless \cite{Blommaert:2018rsf, Anninos:2020hfj, Grewal:2022hlo, Ball:2024hqe}.} For instance, when a gauge or gravity theory is restricted to a subregion, there are physical gauge transformations supported on the boundary of that subregion \cite{Donnelly:2016auv, Freidel:2020xyx, Freidel:2020svx, Freidel:2020ayo}. Another example arises in the study of entanglement entropy across a subregion, where contributions associated with electric flux through the entangling surface are found \cite{Donnelly:2014fua, Donnelly:2015hxa, David:2022jfd}. Furthermore, when the boundary in question is a (stretched) horizon, discrepancies can arise between free energies or entropies computed using Euclidean and Lorentzian techniques \cite{Dowker:2010bu,Casini:2011kv,Anninos:2020hfj,David:2021wrw}. An emerging understanding suggests that these seemingly distinct phenomena are manifestations of the same underlying physics associated with edge modes. In particular the case of Maxwell theory, i.e. pure abelian gauge theory, is now thoroughly understood \cite{Kabat:1995eq,Donnelly:2011hn, Donnelly:2012st, Eling:2013aqa, Radicevic:2014kqa, Donnelly:2014gva, Donnelly:2014fua, Huang:2014pfa, Ghosh:2015iwa, Hung:2015fla, Aoki:2015bsa, Donnelly:2015hxa, Radicevic:2015sza, Pretko:2015zva, Soni:2015yga, Donnelly:2016auv, Zuo:2016knh, Soni:2016ogt, Delcamp:2016eya, Agarwal:2016cir, Blommaert:2018rsf, Blommaert:2018oue, Freidel:2018fsk,Ball:2024hqe}. 

In our previous work \cite{Ball:2024hqe}, we presented a dynamical framework for describing edge modes in Maxwell theory in the presence of a timelike boundary. Central to this framework is the introduction of what we call the dynamical edge mode (DEM) boundary condition, allowing for physical large gauge transformations and their symplectic conjugates. The goal of this paper is to extend this analysis to $p$-form gauge theories for any $p\geq 1$. These theories are intriguing due to their rich geometric structure \cite{10.1007/BFb0075216} and electromagnetic duality, as well as their pivotal roles in various other areas of physics. For instance, the low-energy effective descriptions of string theory and M-theory—namely, 10D and 11D supergravity—feature several massless $p$-form gauge fields. In fact, the relevance of $p$-forms to entanglement entropy in string theory was part of the motivation for this work \cite{Susskind:1994sm, Dabholkar:1994ai, He:2014gva, Mertens:2015adr, Mertens:2016tqv,  Balasubramanian:2018axm,Witten:2018xfj, Donnelly:2020teo, Jiang:2020cqo, Dabholkar:2023tzd, Dabholkar:2024neq, Blommaert:2024cal}.
Additionally, $p$-form gauge theories provide quintessential examples of higher-form symmetries \cite{Gaiotto:2014kfa}. And in even spacetime dimensions $D$, a $p$-form gauge theory with $p = \frac{D-2}{2}$ is conformally invariant, thereby providing an infinite family of CFTs. It is also worth noting that edge modes have been studied in $p$-form BF theories \cite{Fliss:2023dze,Fliss:2023uiv}.

Despite their potential roles in the aforementioned topics, edge modes in the case of $p > 1$ have not been as thoroughly studied as their Maxwell ($p = 1$) counterpart. Nonetheless, some calculations of partition functions and entanglement entropy \cite{Donnelly:2016mlc, Dowker:2017flz, Moitra:2018lxn, Anninos:2020hfj, David:2021wrw, Mukherjee:2023ihb, Dowker:2024cwy} suggest the existence of edge contributions from a $(p-1)$-form residing on the codimension-2 entangling surface, naturally generalizing the $p=1$ case. As we shall see, a natural extension of our DEM framework in \cite{Ball:2024hqe} nicely captures the edge contributions, thus providing a dynamical understanding of these results.

\subsection{Overview of paper}

Our methods are substantially more technical than in the Maxwell case, but our results are conceptually clean, so we state them clearly here. In section \ref{sec:review} we review $p$-form gauge theory. Its partition function contains an alternating tower of functional determinants involving $k$-forms of all degrees $0 \le k \le p$. In section \ref{sec:CovPhaseEdge} we show that $p$-form gauge theory on a Lorentzian manifold $M$ with the dynamical edge mode (DEM) boundary condition \eqref{eq:DEMBC}, reproduced here,
\be {\rm DEM:} \qquad n^\mu F_{\mu a_1\dots a_p}|_{\p M} = 0 = A_{ta_1\dots a_{p-1}}|_{\p M}\; , \ee
splits cleanly into bulk and edge parts. Specifically, in section \ref{subsec:PhaseSplit} we study the phase space on a Cauchy slice $\Sigma$, and we show that the degrees of freedom can be decomposed into bulk modes realized as $p$-forms living on $\Sigma$ and edge modes parametrized by $(p-1)$-forms $\alpha$, $E_\perp$ living on $\p\Sigma$. Here $\alpha$ shifts under large gauge transformations and $E_\perp$ is the ``electric" flux through the boundary. The symplectic form splits under this decomposition,
\be \Omega_{\rm DEM} = \Omega_{\rm bulk} + \Omega_{\rm edge} \; , \qquad \Omega_{\rm edge} = \int_{\p\Sigma} \delta \alpha \wedge *_{\p\Sigma} E_\perp \; , \ee
and consequently the phase space factorizes,
\be \Gam_{\rm DEM} = \Gam_{\rm bulk} \times \Gam_{\rm edge} \; . \ee
The bulk part precisely corresponds to the ``magnetic" boundary condition \eqref{eq:PMCBC}. In section \ref{sec:HamSplit} we proceed to show that the Hamiltonian splits along these same lines,
\be H_{\rm DEM} = H_{\rm bulk} + H_{\rm edge} \; , \qquad H_{\rm edge} = \half \int_{\p\Sigma} E_\perp \wedge *_{\p\Sigma} (K^{-1} E_\perp)\; . \ee
Here $K$ is a certain pseudo-differential operator on $\p\Sigma$, defined in \eqref{eq:defK}. In section \ref{sec:horlim} we study the horizon limit, in which the boundary $\p M$ approaches a bifurcate Killing horizon, and we find that $K$ simplifies to a multiple of the $(p-1)$-form Laplacian on $\p\Sigma$,
\be K \propto \Delta_{p-1}^{\p\Sigma} \; . \ee
With these ingredients in hand, in section \ref{sec:quantum} we quantize and study thermal partition functions. The partition function manifestly factorizes into bulk and edge pieces, and in section \ref{sec:EdgeP} we evaluate the edge part, finding in \eqref{eq:Zbedge-final} that it amounts to the reciprocal of a codimension-two $(p-1)$-form partition function on $\p\Sigma$:
\be \bar Z_{\rm edge}(\beta) \equiv \frac{1}{|\mc{G}^{\p\Sigma}|} \Tr_{\rm edge} e^{-\beta H_{\rm edge}}\propto \frac{1}{\mc{Z}_{\rm PI}^{p-1}[\p\Sigma]} \; . \ee
This is one of our main results. We chose to compute the thermal partition function as a canonical trace, but it can equivalently be obtained from the Euclidean path integral. The Lorentzian boundary that closely hugged the horizon Wick rotates to a Euclidean boundary enclosing a small hole. We say a boundary condition is shrinkable if applying it to an infinitesimally small hole is equivalent to having no hole at all, and in section \ref{sec:shrink} we argue that our DEM boundary condition is shrinkable by comparing with literature on sphere partition functions for $p$-form gauge theory. In section \ref{sec:SDnumerical}, we tabulate numerical values for the bulk and edge parts of the partition function for a sphere of radius $R$. In section \ref{sec:ConfCase}, we discuss entanglement entropy in even dimensions $D$ and comment on the conformal case $p=\frac{D-2}{2}$. Finally in section \ref{sec:DW} we take a different approach and discuss how the non-dynamical prescription of \cite{Donnelly:2015hxa} can be generalized to both $p$-form gauge theory and the free massive $p$-form. Both cases require qualitatively new ingredients compared to Maxwell theory.

We also include a few appendices. In appendix \ref{app:DiffConv} we outline our conventions for differential forms. Appendix \ref{app:horiz-lim} contains details of the horizon limit calculation for the operator $K$ appearing in $H_{\rm edge}$. In appendix \ref{app:sphere_PI} we collect and re-derive relevant results from \cite{Anninos:2020hfj,David:2021wrw,Mukherjee:2023ihb} of free massive and massless $p$-form theories on $dS_D$ static patch and its Euclidean counterpart, namely a round sphere $S^D$. 

\section{Review of $p$-form gauge theory}
\label{sec:review}

In this section we review $p$-form gauge theory and its quantization on a $D$-dimensional manifold. We write $A$ for the $p$-form gauge field and $F = dA$ for its $(p+1)$-form field strength. We take the gauge group to be $U(1)$ with fundamental charge $q$. We assume $p \le D-2$ for simplicity.\footnote{The case of $p=D-1$ is qualitatively different; na\"ively it would be dual to a $\tilde p$-form gauge theory with $\tilde p = D - 2 - p = -1$. A well-known example is Maxwell in $D=2$, where the theory depends only on the volume of the manifold, not its specific geometry. This quasi-topological nature extends to cases where $p = D-1$, which make them extremely tractable and a popular starting point for testing new ideas in gauge theory \cite{Alvarez:1997ma, Donnelly:2012st, Donnelly:2014gva}.}

\subsection{Action and symplectic form}

The action on a Lorentzian manifold $M$ is
\be \label{eq:p-action} S = -\half \int_M F \wedge *_M F = -\frac{1}{2(p+1)!} \int_M F_{\mu_1\dots\mu_{p+1}} F^{\mu_1\dots\mu_{p+1}}\; . \ee
In the course of this paper we will use Hodge duals on a few different manifolds, so we include a subscript to avoid ambiguity. In the latter expression we have suppressed the measure $d^Dx \sqrt{-g}$, which is our convention when integrating scalars (as opposed to forms). Varying the Lagrangian $L= -\half F \wedge *_M F $ gives  
\be \label{eq:lagvar} \delta L = (-)^p \delta A \wedge d*_M F - d\left( \delta A \wedge *_M F \right) \; . \ee
The first term is the equation of motion,
\be d*_M F = 0\; , \ee
which generalizes the usual Maxwell equation of motion. From the second term, we see that on shell the action's variation reduces to a boundary term,
\be \label{eq:var-well-def} \delta S|_{\text{on-shell}} = -\int_{\p M} \delta A \wedge *_M F \; . \ee
For the action to be variationally well-defined, we must choose a boundary condition such that this term vanishes. One option is to require the gauge field to vanish on the boundary,
\be \label{eq:PMCBC} {\rm PEC}: \qquad i_{\p M} A = 0\; , \ee
which directly generalizes the perfectly electrically conducting (PEC) boundary condition in Maxwell theory. Here $i_{\p M}$ indicates the pullback to $\p M$. Another option is to require the boundary normal components of the field strength to vanish,
\be {\rm PMC}: \qquad i_{\p M} *F = 0 \qquad \text{or equivalently} \qquad n^\nu F_{\nu\mu_1\dots\mu_p}|_{\p M} = 0\;, \ee
generalizing the perfectly magnetically conducting (PMC) boundary condition in Maxwell theory. Here $n^\mu$ is the outward unit normal vector to $\partial M$. In section \ref{subsec:PhaseSplit} we will introduce a new boundary condition giving rise to edge modes. For the remainder of this subsection we assume only that some boundary condition has been chosen, making the action variationally well-defined. The space of field configurations allowed by a boundary condition is the starting point for constructing phase space, which is defined by taking the subspace of on-shell configurations and quotienting by symplectically degenerate variations \cite{Lee:1990nz, Wald:1993nt, Iyer:1994ys, Iyer:1995kg}.\footnote{If there are topologically nontrivial gauge transformations then one must also quotient by them by hand. The covariant phase space formalism inherently only knows about infinitesimal variations in fields, as opposed to finite jumps corresponding to topologically nontrivial gauge transformations.} From the Lagrangian variation \eqref{eq:lagvar} we read off the (pre-)symplectic potential density
\be \theta = -\delta A \wedge *_M F\; . \ee
The (pre-)symplectic form is obtained by varying $\theta$ and integrating over a Cauchy slice $\Sigma$, giving
\be \Omega = \int_\Sigma \delta A \wedge *_M \delta F \; , \ee
where $\wedge$ refers only to the wedge product on spacetime. The wedge product on phase space is left implicit. Note that on shell the symplectic form is invariant under bulk deformations of $\Sigma$ since the integrand is closed,
\be d\left(\delta A \wedge *_M \delta F\right) = \delta F \wedge *_M \delta F = 0\; . \ee
The first step used the equation of motion, and the last step used antisymmetry of the wedge form on phase space. Invariance of the symplectic form under deformations of the boundary of $\Sigma$ is equivalent to the action being variationally well-defined.

Now we turn our attention to gauge transformations. A shift $\delta A = d\lam$ has no effect on the field strength $F = dA$. Plugging such a variation into the symplectic form gives 
\be \Omega = \int_\Sigma d\lam \wedge *_M \delta F = \int_{\p\Sigma} \lam \wedge *_M \delta F\; , \ee
where we have used the equation of motion. We see that this vanishes for any $\lam$ with vanishing boundary pullback, so such transformations are symplectically trivial and unphysical. We refer to them as small gauge transformations. In contrast, large gauge transformations are those with nontrivial boundary pullback. They are physical symmetries, with corresponding charges
\be Q[\lam] = \int_{\p\Sigma} \lam \wedge *_M F\; . \ee
This is all highly analogous to the story in Maxwell theory.

\subsection{Number of degrees of freedom}
\label{sec:num-dof}

It will be useful later as a sanity check to know the number of local degrees of freedom, or polarizations, of a $p$-form gauge field. It can be deduced as follows. A $p$-form initially has ${D \choose p}$ independent components but, at least locally, we can use gauge transformations to kill the time component, $t^\mu A_{\mu\dots} = 0$. This is called temporal gauge. We still have residual time-independent gauge transformations, and we can use them to further set $v^\mu A_{\mu\dots} = 0$ on shell, for some fixed spatial vector field $v^\mu$, perhaps corresponding to a particular coordinate. This leaves us with
\begin{align}
    \binom{D}{p} - \binom{D-1}{p-1}- \binom{D-2}{p-1}=\binom{D-2}{p}
\end{align}
independent physical polarizations. Note that $p=1$ gives ${D-2 \choose 1} = D-2$ polarizations, which is correct. This quick argument informs us about the number of $D$-dimensional degrees of freedom, but says nothing about the edge modes, which are effectively lower-dimensional degrees of freedom.

\subsection{Euclidean partition function}\label{sec:ZPIgeneral}

Next we consider the $p$-form gauge theory partition function on a closed, connected $D$-dimensional Euclidean manifold $\mc{M}$. See \cite{Donnelly:2016mlc} for a detailed discussion. It will be helpful to first review the $p=1$ case, i.e. Maxwell theory, whose partition function on a closed manifold is given by the path integral \cite{Donnelly:2013tia}
\be \mc{Z}_{\rm PI}^{p=1}[\mc{M}] = \sum_{\rm bundles} \int \frac{\mc{D} A}{|\mc{G}|} \, e^{-S[F]}\; , \ee
where $|\mc{G}|$ is the volume of the gauge group. The sum is over magnetic bundles obeying the Dirac quantization condition, labeled by the second integer cohomology group $H^2(\mc{M}, \integ)$. The field strength in any bundle can be written globally as
\be F = \mathscr{F} + dA\; , \ee
with $\mathscr{F} \in \frac{2\pi}{q} \mc{H}^2(\mc{M}, \integ)$, where $\mc{H}^2(\mc{M}, \integ)$ is the subset of harmonic two-forms whose integral over any closed submanifold gives an integer. Thus we can always treat $A$ as residing in the trivial bundle (and thus being globally well-defined). Alternatively one could split into patches and write locally $F = dA$ with appropriate transition functions between the patches. With the choice here the action splits as $S[F] = S[\mathscr{F}] + S[dA]$ and the partition function factorizes as
\be \mc{Z}_{\rm PI}^{p=1}[\mc{M}] = \sum_{\mathscr{F}\in \frac{2\pi}{q} \mc{H}^2(\mc{M},\integ)} e^{-S[\mathscr{F}]} \times \int \frac{\mc{D} A}{|\mc{G}|} \, e^{-S[dA]}\; . \label{eq:p1fullZPI}\ee
The measure $\mc{D} A$ integrates over all one-forms on $\mc{M}$. The gauge group $\mc{G}$ consists of all $U(1)$-valued functions on $\mc{M}$, acting by $A \to A + d\phi$. The periodicity is $\phi \sim \phi + \frac{2\pi}{q}$. Note that $\phi$'s differential $d\phi$ is real-valued and constitutes a one-form in the standard sense. However, it is not necessarily exact due to the periodicity of $\phi$. Rather $d\phi$'s cohomology class is in the first integer cohomology group $\frac{2\pi}{q} H^1(\mc{M}, \integ)$. This perspective, of gauge transformations as representatives of an integer cohomology group, turns out to generalize to $p$-forms more easily than the perspective of gauge transformations as multi-valued functions.

To proceed with the evaluation of the path integral it helps to Hodge decompose $A$ into exact, harmonic, and co-exact parts,
\be A = A_{\rm ex} + A_{\rm harm} + A_{\rm co}\; , \qquad \mc{D} A = \mc{D} A_{\rm ex} \mc{D} A_{\rm harm} \mc{D} A_{\rm co}\; . \ee
We can do something similar for the gauge parameter $\phi$, although it has an additional topological part and no exact part. Also its harmonic part is simply the constant function. We have
\be \phi = \phi_{\rm topo} + \phi_{\rm const} + \phi_{\rm co} \ee
with $d\phi_{\rm topo} \in \frac{2\pi}{q} \mc{H}^1(\mc{M}, \integ)$. We therefore have\footnote{Technically for the topological part we should speak in terms of quotienting by its group action, rather than dividing by its ``volume". Nevertheless we write $|\mc{G}_{\rm topo}|$ to keep the notation compact.}
\begin{align}
     |\mc{G}| = |\mc{G}_{\rm topo}| \left(\int \mc{D} \phi_{\rm const} \right) \left(\int \mc{D} \phi_{\rm co}\right) \; .
\end{align}
Plugging these into the path integral \eqref{eq:p1fullZPI}, only the $A_{\rm co}$ part affects the field strength, and therefore the action. The rest, namely $A_{\rm ex}$ and $A_{\rm harm}$, get integrated over freely. We would run into infinities if not for several cancellations. First, by definition we have
\be A_{\rm ex} = d\chi \ee
for some function $\chi$. We remove the zero mode ambiguity by requiring $\chi$ to integrate to zero. We get a Jacobian determinant $\mc{D} A_{\rm ex} = \mc{D}(d\chi) = \det'\left(\Delta_0\right)^{\frac12} \mc{D}'\chi$, where $\Delta_0$ is the Laplacian on zero-forms and the prime(s) indicate the omission of the zero mode. We then have the cancellation
\be \frac{\int \mc{D}'\chi}{\int\mc{D}\phi_{\rm co}} = 1\; . \ee
The integration over harmonic one-forms $A_{\rm harm}$ gives the volume of $H^1(\mc{M}, \mathbb{R})$, which is infinite if it is nontrivial, but the quotient by $\mc{G}_{\rm topo} \cong \frac{2\pi}{q} H^1(\mc{M}, \integ)$ renders it finite. Finally the integral $|U(1)| \equiv \int \mc{D} \phi_{\rm const}$ over the constant gauge transformation gives the volume of $U(1)$ with a certain measure. Overall we have
\be \mc{Z}_{\rm PI}^{p=1}[\mc{M}] = \sum_{\mathscr{F}\in \frac{2\pi}{q} \mc{H}^2(\mc{M},\integ)} e^{-S[\mathscr{F}]} \det\hspace{0mm}'\left(\Delta_0\right)^{\frac12} \left| \frac{H^1(\mc{M}, \mathbb{R})}{\frac{2\pi}{q} H^1(\mc{M}, \integ)} \right| \frac{1}{|U(1)|} \int \mc{D} A_{\rm co} \, e^{-S[dA_{\rm co}]}\; . \ee
Noting that $d*_\mc{M} A_{\rm co} = 0$ by definition, the (Euclidean) action reduces to
\be S[dA_{\rm co}] = \half \int_\mc{M} dA_{\rm co} \wedge *_\mc{M} dA_{\rm co} = \half \int_\mc{M} A_{\rm co} \wedge *_\mc{M} \Delta_1 A_{\rm co} \ee
where $\Delta_1 = -(-)^D *_\mc{M} d*_\mc{M} d - d*_\mc{M} d*_\mc{M}$ is the Hodge Laplacian on one-forms. The path integral over $A_{\rm co}$ will then give the determinant of the Laplacian restricted to act on co-closed, i.e. divergenceless or transverse, one-forms. Then 
\be \mc{Z}_{\rm PI}^{p=1}[\mc{M}] = \frac{\det\hspace{0mm}'\left(\Delta_0\right)^{\frac12}}{\det\hspace{0mm}'\left(\Delta_1^T\right)^{\frac12}} \, \frac{1}{|U(1)|} \left| \frac{H^1(\mc{M}, \mathbb{R})}{\frac{2\pi}{q} H^1(\mc{M}, \integ)} \right| \sum_{\mathscr{F}\in \frac{2\pi}{q} \mc{H}^2(\mc{M},\integ)} e^{-S[\mathscr{F}]}\; . \ee
The superscript $T$ in $\Delta_1^T$ indicates that the determinant is only over the space of ``transverse", i.e. divergenceless, one-forms. To actually evaluate the functional determinants and volumes here one should choose some explicit path integral measure. We do this for $|U(1)|$ below in \eqref{eq:U(1)vol}, for general $p$. When a homology group is trivial, the factor $\left| \frac{H^k(\mc{M}, \mathbb{R})}{\frac{2\pi}{q} H^k(\mc{M}, \integ)} \right|$ reduces to unity.

\paragraph{$p=2$ and beyond}

The situation for $p$-forms is similar, but with the key difference that there is an alternating tower of gauge transformations rather than a single gauge group, and their relative Jacobians lead to an alternating tower of determinants. This can be realized in terms of ghosts-for-ghosts \cite{Townsend:1979hd, Siegel:1980jj, Copeland:1984qk}, but we will give a strictly geometric description \cite{Obukhov:1982dt}. Let us warm up with the $p=2$ case,
\be \mc{Z}_{\rm PI}^{p=2}[\mc{M}] = \sum_{\rm bundles} \frac{\mc{D} A}{|\mc{G}_1/\mc{G}_0|} \, e^{-S[F]} \; . \ee
The sum is over ``magnetic" bundles obeying an analogue of Dirac quantization \cite{Teitelboim:1985yc}, labeled by $H^3(\mc{M}, \integ)$. Once again we can globally split the field strength into harmonic and exact parts,
\be F = \mathscr{F} + dA \; . \ee
Here $\mathscr{F} \in \frac{2\pi}{q} \mc{H}^3(\mc{M}, \integ)$ and $A$ is an arbitrary globally defined two-form (i.e. it resides in the trivial bundle). The group $\mc{G}_1$ shifts $A$ by a representative of $\frac{2\pi}{q} H^2(\mc{M}, \integ)$, which can be split into a $\frac{2\pi}{q} \mc{H}^2(\mc{M}, \integ)$ part and an exact part $d\lam$, with $\lam$ an arbitrary one-form. (This includes closed $\lam$.) The group $\mc{G}_0$ acts as $\lam \to \lam + d\phi$ with $\phi$ an arbitrary $U(1)$-valued function, just like the gauge group in the $p=1$ case. After Hodge decomposing the respective forms appearing, we have
\begin{align}
\mc{D} A &= \mc{D} A_{\rm ex} \mc{D} A_{\rm harm} \mc{D} A_{\rm co}\nn\\
     |\mc{G}_1| & = \left| \frac{2\pi}{q} H^2(\mc{M}, \integ)\right| \left(\int \mc{D} \lam_{\rm ex} \right) \left( \int\mc{D} \lam_{\rm harm} \right) \left( \int\mc{D} \lam_{\rm co}\right) \nn\\
     |\mc{G}_0| & = \left| \frac{2\pi}{q} H^1(\mc{M}, \integ)\right| \left(\int \mc{D} \phi_{\rm const} \right) \left( \int\mc{D} \phi_{\rm co}\right)\;. 
\end{align}
As before, only the integration over $A_{\rm co}$ is weighted by the action. The integrations over harmonic forms are rendered finite by quotients
\begin{align}
    \frac{\int \mc{D} A_{\rm harm} }{\left| \frac{2\pi}{q} H^2(\mc{M}, \integ)\right|}  = \left| \frac{H^2(\mc{M}, \mathbb{R})}{\frac{2\pi}{q} H^2(\mc{M}, \integ)} \right| \; , \qquad 
        \frac{\int\mc{D} \lam_{\rm harm}}{\left| \frac{2\pi}{q} H^1(\mc{M}, \integ)\right|}&= \left| \frac{H^1(\mc{M}, \mathbb{R})}{\frac{2\pi}{q} H^1(\mc{M}, \integ)} \right| \; ,
\end{align}
while the integration of the exact part of a $k$-form naturally cancels against the co-exact part of a $(k-1)$-form up to a Jacobian factor
\begin{align}
    \frac{\int \mc{D} A_{\rm ex}}{\int\mc{D} \lam_{\rm co}} = \det\hspace{0mm}'\left(\Delta_1^T\right)^{\frac12} \; , \qquad \frac{\int \mc{D} \lam_{\rm ex} }{\int\mc{D} \phi_{\rm co}}  = \det\hspace{0mm}'\left(\Delta_0\right)^{\frac12} \; .  
\end{align}
Putting these together, the partition function becomes
\begin{align}
    & \quad \mc{Z}_{\rm PI}^{p=2}[\mc{M}]  \nn\\
& = \hspace{-2mm} \sum_{\mathscr{F}\in\frac{2\pi}{q}\mc{H}^3(\mc{M},\integ)} \hspace{-5mm} e^{-S[\mathscr{F}]} \det\hspace{0mm}'\left(\Delta_1^T\right)^{\frac12} \left| \frac{H^2(\mc{M}, \mathbb{R})}{\frac{2\pi}{q} H^2(\mc{M}, \integ)} \right| \det\hspace{0mm}'\left(\Delta_0\right)^{-\frac12} \left| \frac{H^1(\mc{M}, \mathbb{R})}{\frac{2\pi}{q} H^1(\mc{M}, \integ)} \right|^{-1} |U(1)| \int \mc{D} A_{\rm co} \, e^{-S[dA_{\rm co}]} \; .
\end{align}
The remaining path integral gives $\det'\left(\Delta_2^T\right)^{-\frac12}$, yielding the final answer
\begin{align}
    &\quad \mc{Z}_{\rm PI}^{p=2}[\mc{M}] \nn\\
& = \hspace{-2mm} \sum_{\mathscr{F}\in\frac{2\pi}{q}\mc{H}^3(\mc{M},\integ)} \hspace{-5mm} e^{-S[\mathscr{F}]} \det\hspace{0mm}'\left(\Delta_2^T\right)^{-\frac12} \left| \frac{H^2(\mc{M}, \mathbb{R})}{\frac{2\pi}{q} H^2(\mc{M}, \integ)} \right| \det\hspace{0mm}'\left(\Delta_1^T\right)^{\frac12} \left| \frac{H^1(\mc{M}, \mathbb{R})}{\frac{2\pi}{q} H^1(\mc{M}, \integ)} \right|^{-1} \det\hspace{0mm}'\left(\Delta_0\right)^{-\frac12} |U(1)| \; . 
\end{align}
The pattern for general $p$ has begun to emerge. Using the same methods just outlined, one can show
\be \ba \mc{Z}_{\rm PI}^p[\mc{M}] & = \sum_{\mathscr{F}\in\frac{2\pi}{q} \mc{H}^{p+1}(\mc{M}, \integ)} \hspace{-5mm} e^{-S[\mathscr{F}]} \int \frac{\mc{D} A}{|\mc{G}_{p-1}/(\mc{G}_{p-2}/\dots)|} \, e^{-S[dA]} \\
& = \sum_{\mathscr{F}\in\frac{2\pi}{q} \mc{H}^{p+1}(\mc{M}, \integ)} \hspace{-5mm} e^{-S[\mathscr{F}]} \prod_{k=0}^p \left( \det\hspace{0mm}'\left(\Delta_k^T\right)^{-\frac12} \left| \frac{H^k(\mc{M}, \mathbb{R})}{\frac{2\pi}{q} H^k(\mc{M}, \integ)} \right| \right)^{(-)^{p-k}} \; . \ea \ee
Note this uses $|U(1)| = \frac{H^0(\mc{M}, \mathbb{R})}{\frac{2\pi}{q} H^0(\mc{M}, \integ)}$. The nested quotient of gauge groups can be viewed as a single gauge group with a particular measure,
\be \label{eq:fullG} \ba |\mc{G}| & \equiv |\mc{G}_{p-1} / (\mc{G}_{p-2} / \dots)| \\
& = |\mc{G}_{\rm nontriv}| \left| \frac{H^{p-1}(\mc{M}, \mathbb{R})}{\frac{2\pi}{q} H^{p-1}(\mc{M}, \integ)} \right| \, \prod_{k=0}^{p-2} \left( \det\hspace{0mm}'\left(\Delta_k^T\right)^{-\frac12} \left| \frac{H^k(\mc{M}, \mathbb{R})}{\frac{2\pi}{q} H^k(\mc{M}, \integ)} \right| \right)^{(-)^{p-k+1}} \ea \ee
where $\mc{G}_{\rm nontriv}$ is the nontrivial part of $\mc{G}_{p-1}$, i.e. $\frac{2\pi}{q} \mc{H}^p(\mc{M},\integ)$ plus $d\lam$ with $\lam$ a non-closed $(p-1)$-form. Then the measure can be written simply as $\mc{D}A / |\mc{G}|$. For later use we evaluate $|U(1)|$ explicitly here. A natural measure for the scalar gauge parameter $\phi$ is $\mc{D}\phi = \prod_n \frac{\mu^{p+1} d\phi_n}{\sqrt{2\pi}}$, associated with the Laplacian eigendecomposition $\phi(x) = \sum_n \phi_n f_n(x)$. Here the arbitrary mass scale $\mu$ is analogous to a UV cutoff, and is included so that the measure is dimensionless. The normalized zero mode is $f_0(x) = V_{\mc M}^{-1/2}$, where $V_{\mc M}$ is the volume of $\mc{M}$. The zero mode coefficient then inherits the periodicity $\phi_0 \sim \phi_0 + \frac{2\pi}{q} V_{\mc M}^{1/2}$. We then have
\be \label{eq:U(1)vol} |U(1)| \equiv \int_0^{\frac{2\pi}{q} \sqrt{V_\mc{M}}} \frac{\mu^{p+1}}{\sqrt{2\pi}} d\phi_0 = \frac{\mu^{p+1}}{q} \sqrt{2\pi V_\mc{M}} \; . \ee
Some authors write the partition function in terms of full Laplacian determinants $\det'\left(\Delta_k\right)$ as opposed to the transverse determinants used here. The relationship between the two formulas follows recursively from the identity
\be \label{eq:detid} \det\hspace{0mm}'\left(\Delta_k\right) = \det\hspace{0mm}'\left(\Delta_k^T\right) \det\hspace{0mm}'\left(\Delta_{k-1}^T\right)\; . \ee
The functional determinant factors tell us about the local degrees of freedom of the theory. We now carry out a sanity check to confirm that in total there are ${D-2 \choose p}$ local degrees of freedom, as argued above. A $k$-form has ${D \choose k}$ independent components, so a determinant $\det'\left(\Delta_k\right)^{-\frac12}$ represents ${D \choose k}$ degrees of freedom. Similarly its reciprocal represents $-{D \choose k}$ degrees of freedom. Let us write the number of degrees of freedom in a transverse $k$-form determinant as $n_k$. From \eqref{eq:detid} we apparently have $n_k + n_{k-1} = {D \choose k}$. This recursion relation is solved by
\be n_k = {D-1 \choose k}\;. \ee
The total number of degrees of freedom in the partition function $\mc{Z}_{\rm PI}^p[\mc{M}]$ is then
\be \sum_{k=0}^p (-)^{p-k} {D-1 \choose k} = {D-2 \choose p}\; , \ee
showing that the somewhat abstruse tower of determinants is nevertheless consistent with the more intuitive gauge-fixing argument in section \ref{sec:num-dof}.

\section{Bulk and edge modes in $p$-form gauge fields}
\label{sec:CovPhaseEdge}

In this section we restrict attention to a static Lorentzian manifold $M$. We take the metric to be
\be \label{eq:static-metric} ds^2 = g_{tt} dt^2 + g_{ij} dx^i dx^j \ee
with all components independent of time. We denote a surface of constant $t$ by $\Sigma$. We further introduce coordinates $x^a$ on $\p\Sigma$, so that the metric on $\p M$ can be written as
\be ds^2|_{\p M} = g_{tt} dt^2 + g_{ab} dx^a dx^b\;. \ee
The vectors $\p_a$ are by definition orthogonal to $\p_t$ and the outward unit boundary normal vector $n^\mu$. Note $n^\mu$ can also be viewed as the outward normal to $\p\Sigma$ in $\Sigma$.

\subsection{Phase space split}
\label{subsec:PhaseSplit}

We now introduce the dynamical edge mode (DEM) boundary condition,
\be {\rm DEM:} \qquad n^\mu F_{\mu a_1\dots a_p}|_{\p M} = 0 = A_{ta_1\dots a_{p-1}}|_{\p M}\; . \label{eq:DEMBC} \ee
This is a direct generalization of the DEM boundary condition for Maxwell theory, $n^\mu F_{\mu a}|_{\p M} = 0 = A_t|_{\p M}$, introduced in \cite{Ball:2024hqe}. One can see by plugging into \eqref{eq:var-well-def} that it makes the action variationally well-defined. Let us now analyze its phase space, defined as the space of solutions quotiented by the symplectically trivial variations.\footnote{We also quotient by any topologically nontrivial gauge transformations by hand.} First, recall the symplectic form
\be \Omega = \int_\Sigma \delta A \wedge *_M \delta F\; . \ee
Since we are restricting to solutions, it is independent of the choice of time slice $\Sigma$. Fix some choice. The symplectic form manifestly depends only on the fields $i_\Sigma A$, $i_\Sigma {*_M F}$ on $\Sigma$, so these must be sufficient to parametrize phase space (with some redundancy). We find it convenient to give a name to the pullback of the dual field strength,
\be E_{i_1\dots i_p} \equiv -\sqrt{-g^{tt}} F_{ti_1\dots i_p} \; . \ee
This $E$ is analogous to the usual electric field. The symplectic form now reads
\be \label{eq:symform} \Omega = \int_\Sigma \delta A \wedge *_\Sigma \delta E\; , \ee
and henceforth we think of the quantities $A$, $E$ as living only on $\Sigma$, essentially forgetting about $M$. The choices of $A$ and $E$ on $\Sigma$ are completely independent from each other. The equation of motion implies a constraint on $E$, namely
\be d*_\Sigma E = i_\Sigma d{*_M F} = 0\; . \ee
This generalizes the Gauss constraint. As a consequence there is some gauge redundancy in $A$. As noted above, taking $\delta A = d\lam$ leads to
\be \label{eq:Qvar} \Omega = \int_\Sigma d\lam \wedge *_\Sigma \delta E = \int_{\p\Sigma} \lam \wedge *_\Sigma \delta E\; . \ee
This vanishes whenever $\lam$'s boundary pullback does, i.e. when $\lam$ is a small gauge transformation, but if $i_{\p\Sigma} \lam \ne 0$ then this can be nonzero because the DEM boundary condition allows nonzero flux $i_{\p\Sigma} *_\Sigma E$ through the boundary. We give a name to this flux, which is a $(p-1)$-form on $\p\Sigma$, defining
\be E_{\perp,a_1\dots a_{p-1}} \equiv n^i E_{i a_1\dots a_{p-1}} \; . \ee
The charge variation \eqref{eq:Qvar} then reduces to $\int_{\p\Sigma} \lam \wedge *_{\p\Sigma} \delta E_\perp$, which directly generalizes the Maxwell expression $\int_{\p\Sigma} \lam \delta E_\perp$ where both $\lam$, $E_\perp$ were scalars.

At this point it is natural to introduce a generalization of the Hodge decomposition for manifolds with boundary, called the Hodge-Morrey-Friedrichs decomposition \cite{Schwarz}. Explicitly, it states that the space of $k$-forms on $\Sigma$ splits as
\be \Omega^k(\Sigma) = \mc{E}^k(\Sigma) \oplus \mc{H}^k(\Sigma) \oplus \mc{C}^k(\Sigma) \; . \ee
Here $\mc{E}^k(\Sigma)$ is the space of exact $k$-forms on $\Sigma$ with vanishing pullback to $\p\Sigma$, $\mc{H}^k(\Sigma)$ is the space of ``strongly" harmonic $k$-forms $\omega$ satisfying $d\omega = *_\Sigma d *_\Sigma \omega = 0$, and $\mc{C}^k(\Sigma)$ is the space of co-exact $k$-forms on $\Sigma$ whose Hodge duals have vanishing pullback to $\p\Sigma$. We will refer to forms annihilated by the Laplacian as weakly harmonic. On a boundaryless manifold this is equivalent to being strongly harmonic, but that is not the case here. Another difference from the boundaryless case is that $\mc{H}^k(\Sigma)$ is infinite-dimensional. The Hodge-Morrey-Friedrichs decomposition also provides the refinement
\be \mc{H}^k(\Sigma) = \mc{H}^k_{\rm ex}(\Sigma) \oplus \mc{H}^k_N(\Sigma) \ee
where $\mc{H}^k_{\rm ex}(\Sigma)$ is the subspace of exact strongly harmonic forms, and $\mc{H}^k_N(\Sigma)$ is the subspace of strongly harmonic forms whose Hodge duals have vanishing pullback to $\p\Sigma$. We will also define
\be \mc{C}^k_{\rm cc}(\Sigma) \equiv \mc{H}^k_N(\Sigma) \oplus \mc{C}^k(\Sigma) \; , \ee
which is the space of all co-closed forms whose Hodge duals have vanishing pullback to $\p\Sigma$. The decomposition of most relevance to us is then
\be \Omega^k(\Sigma) = \mc{E}^k(\Sigma) \oplus \mc{H}^k_{\rm ex}(\Sigma) \oplus \mc{C}^k_{\rm cc}(\Sigma) \; . \ee
In fact we will actually use a twisted version of this decomposition. Hodge theory is usually formulated in terms of the exterior derivative $d$ and its adjoint $d^\dag = (-)^{1+(D-1)(p-1)} *_\Sigma d*_\Sigma$ with respect to the natural inner product on forms on $\Sigma$, but it generalizes to any elliptic complex. We will use a version of Hodge theory based on
\be \label{eq:twistdiff} d \; , \qquad d^\ddagger = \frac{1}{S} d^\dag S \ee
for some positive function $S$, where the modified adjoint $\ddagger$ corresponds to the rescaled inner product
\be (\omega, \psi) = \int_\Sigma S \, \omega \wedge *_\Sigma \psi \; , \qquad \omega, \psi \in \Omega^k(\Sigma) \; . \ee
The factor of $S$ is related to the choice of time evolution. We will always use
\be S = \sqrt{-g^{tt}}, \ee
corresponding to static time $t$. We will refer to forms annihilated by $d^\ddagger$ as $S$-co-closed, and forms annihilated by the $S$-Laplacian $\Delta \equiv (d + d^\ddagger)^2$ as weakly $S$-harmonic. This leads to new Hodge-Morrey-Friedrichs decompositions
\be \Omega^k(\Sigma) = \mc{E}'^k(\Sigma) \oplus \mc{H}'(\Sigma) \oplus \mc{C}'^k(\Sigma) \ee
and
\be \Omega^k(\Sigma) = \mc{E}'^k(\Sigma) \oplus \mc{H}'^k_{\rm ex}(\Sigma) \oplus \mc{C}'^k_{\rm cc}(\Sigma) \ee
where now $\mc{H}'^k(\Sigma)$, $\mc{C}'^k(\Sigma)$, $\mc{H}'^k_{\rm ex}(\Sigma)$, and $\mc{C}'^k_{\rm cc}(\Sigma)$ are $S$-co-closed. It is not hard to see that
\be \mc{E}'^k(\Sigma) = \mc{E}^k(\Sigma) \; , \qquad \mc{C}'^k(\Sigma) = S^{-1} \mc{C}^k(\Sigma) \; , \qquad \mc{C}'^k_{\rm cc}(\Sigma) = S^{-1} \mc{C}^k_{\rm cc}(\Sigma) \; , \ee
but the relationships between the primed and unprimed harmonic spaces are more subtle. To bring these tools to bear on our phase space, let us finally write 
\be \Omega^k(\Sigma) = \mc{E}^k(\Sigma) \oplus \mc{H}'^k_{\rm ex}(\Sigma) \oplus S^{-1} \mc{C}^k_{\rm cc}(\Sigma) \; . \ee
Applied to our field $A \in \Omega^p(\Sigma)$, we see that $\mc{E}^p(\Sigma)$ is precisely the part of $A$ affected by small gauge transformations. We therefore choose to set it to zero, leaving no further small gauge redundancy. This can also be understood as imposing a rescaled version of Coulomb gauge,
\be d*_\Sigma S A = 0 \; . \ee
We write the remaining field as
\be A = \tilde A + d\alpha \ee
with\footnote{Since $\tilde A$ satisfies the boundary condition $n^\mu F_{\mu a_1\dots a_p}|_{\p\Sigma}=0$ it is not an arbitrary element of $S^{-1} \mc{C}^p_{\rm cc}(\Sigma)$. In contrast $d\alpha \in \mc{H}'^p_{\rm ex}(\Sigma)$ is arbitrary.}
\be \tilde A \in S^{-1} \mc{C}^p_{\rm cc}(\Sigma) \; , \qquad d\alpha \in \mc{H}'^p_{\rm ex}(\Sigma) \; . \ee
There is ambiguity in shifting $\alpha$ by any closed form. We can eliminate this by choosing $\alpha \in S^{-1} \mc{C}^{p-1}(\Sigma)$. Since $d^\ddagger d\alpha = d^\ddagger \alpha = 0$ it is in particular weakly $S$-harmonic, $\Delta\alpha = 0$, and so it can be recovered from its (Dirichlet) boundary data $i_{\p\Sigma}\alpha$ and $i_{\p\Sigma} {*_\Sigma \alpha} = 0$. Next we apply our decomposition to the rescaled electric field $S^{-1} E$. The Gauss constraint $0 = d*_\Sigma E = d*_\Sigma S (S^{-1} E)$ implies that $S^{-1} E$ has no part in $\mc{E}^k(\Sigma)$, and therefore we can write
\be \label{eq:Esplit} E = \tilde E + S d\beta \ee
with
\be \tilde E \in \mc{C}^p_{\rm cc}(\Sigma) \; , \qquad d\beta \in \mc{H}'^p_{\rm ex}(\Sigma) \; . \ee
We eliminate the ambiguity in $\beta$ by choosing $\beta \in S^{-1} \mc{C}^{p-1}(\Sigma)$. Just like for $\alpha$, we note $d^\ddagger d\beta = d^\ddagger \beta = 0$ and so $\beta$ is weakly $S$-harmonic and it can be recovered from its (Dirichlet) boundary data $i_{\p\Sigma} \beta$ and $i_{\p\Sigma} {*_\Sigma \beta} = 0$. Similarly note that $E_\perp$, along with $i_{\p\Sigma} *_\Sigma \beta = 0$, amounts to Neumann data\footnote{In general this type of data comes with integrability conditions on $E_\perp$ and ambiguities in $\beta$ \cite{Schwarz}, but the co-closedness of $E$ and the choice $\beta \in S^{-1} \mc{C}^{p-1}(\Sigma)$ resolve these subtleties.} for $\beta$,
\be *_{\p\Sigma} E_\perp = i_{\p\Sigma} *_\Sigma S d\beta \; . \ee
Plugging these decompositions into the symplectic form \eqref{eq:symform}, we get
\be \ba \Omega & = \int_\Sigma \left(\delta\tilde A + d\delta\alpha \right) \wedge *_\Sigma \left(\delta\tilde E + S d\delta\beta \right) \\
& = \int_\Sigma \delta\tilde A \wedge *_\Sigma \delta\tilde E + \int_{\p\Sigma} \delta\alpha \wedge *_\Sigma S d\delta\beta \\
& = \int_\Sigma \delta\tilde A \wedge *_\Sigma \delta\tilde E + \int_{\p\Sigma} \delta\alpha \wedge *_{\p\Sigma} \delta E_\perp \\
& \equiv \Omega_{\rm bulk} + \Omega_{\rm edge} \; . \ea \ee
The bulk phase space variables $\tilde A$, $\tilde E$ and the edge phase space variables $i_{\p\Sigma} \alpha$, $E_\perp$ are completely independent of each other, so the split of integrals here shows that the phase space factorizes into
\be \label{eq:phase-fac} \Gam_{\rm DEM} = \Gam_{\rm bulk} \times \Gam_{\rm edge} \; . \ee
Furthermore $\tilde A, \tilde E$ precisely correspond to the PMC boundary condition, so we can identify
\be \Gam_{\rm bulk} = \Gam_{\rm PMC} \; . \ee
This is a direct generalization of the Maxwell, i.e. $p=1$, case.

\subsection{Hamiltonian split}
\label{sec:HamSplit}

Next we study the Hamiltonian and find a similar split. First recall our coordinates $(t, x^i)$ from \eqref{eq:static-metric} and break the (Lorentzian) Lagrangian into electric and magnetic parts as
\be L = -\frac{\sqrt{-g}}{2(p+1)!} F_{\mu_1\dots\mu_{p+1}} F^{\mu_1\dots\mu_{p+1}} = \frac{\sqrt{-g}}{2(p+1)!} \left( (p+1) E_{i_1\dots i_p} E^{i_1\dots i_p} - F_{i_1\dots i_{p+1}} F^{i_1\dots i_{p+1}} \right)\; . \ee
The Hamiltonian density differs only by flipping the sign of the magnetic part,
\be H_{\rm density} = \frac{\sqrt{-g}}{2(p+1)!} \left( (p+1) E_{i_1\dots i_p} E^{i_1\dots i_p} + F_{i_1\dots i_{p+1}} F^{i_1\dots i_{p+1}} \right)\; . \ee
Restricting to $\Sigma$, we can view this as a function of our phase space variables $A$ and $E$. The magnetic part depends only on $A$ since it involves only spatial derivatives, and in fact $\alpha$ drops out. Since it depends only on $\tilde A$, we will write it as $\tilde F_{i_1\dots i_{p+1}}$. The full Hamiltonian generating time $\p_t$ is obtained by integrating over a slice, 
\be H = \int_\Sigma d^{D-1}x \, H_{\rm density} \; . \ee
Defining
\be H_{\rm bulk} \equiv \int_\Sigma d^{D-1}x \frac{\sqrt{-g}}{2(p+1)!} \left( (p+1) \tilde E_{i_1\dots i_p} \tilde E^{i_1\dots i_p} + \tilde F_{i_1\dots i_{p+1}} \tilde F^{i_1\dots i_{p+1}} \right) \ee
we have
\be \label{eq:Ham-split} \ba H & = H_{\rm bulk} + \half \int_\Sigma \sqrt{-g_{tt}} (2 S d\beta \wedge *_\Sigma \tilde E + S d\beta \wedge *_\Sigma S d\beta) \\
& = H_{\rm bulk} + \half \int_\Sigma (2d\beta \wedge *_\Sigma \tilde E + d\beta \wedge *_\Sigma S d\beta) \\
& = H_{\rm bulk} + \half \int_\Sigma d\beta \wedge *_\Sigma S d\beta \\
& = H_{\rm bulk} + \half \int_{\p\Sigma} \beta \wedge *_\Sigma S d\beta \\
& = H_{\rm bulk} + \half \int_{\p\Sigma} \beta \wedge *_{\p\Sigma} E_\perp \\
& = H_{\rm bulk} + H_{\rm edge} \ea \ee
where in the second line we used $\sqrt{-g_{tt}} = S^{-1}$, in the third we integrated by parts and used $d*_\Sigma \tilde E = 0 = i_{\p\Sigma} *_\Sigma \tilde E$, and in the fourth we used $d*_\Sigma Sd\beta = 0$. The form of the edge Hamiltonian $H_{\rm edge}$ motivates defining an operator $K$ such that
\be \label{eq:defK} K i_{\p\Sigma} \beta = E_\perp \; . \ee
It can be understood as follows. Since $\beta$ is weakly $S$-harmonic, it is determined by its boundary data. Part of this data is $i_{\p\Sigma} *_\Sigma \beta = 0$, and for the other part we can take either ``Dirichlet" data $i_{\p\Sigma} \beta$ or ``Neumann" data $E_\perp$ with $*_{\p\Sigma} E_\perp = S i_{\p\Sigma} *_\Sigma d\beta$ \cite{Schwarz}. We then define the Dirichlet-to-Neumann operator $K: \Omega^{p-1}(\p\Sigma) \to \Omega^{p-1}(\p\Sigma)$ as the map from Dirichlet data to the corresponding Neumann data. This is once again a direct generalization of the Maxwell case. In terms of $K$ we can write
\be H_{\rm edge} = \half \int_{\p\Sigma} \beta \wedge *_{\p\Sigma} E_\perp = \half \int_{\p\Sigma} \beta \wedge *_{\p\Sigma} K \beta = \half \int_{\p\Sigma} E_\perp \wedge *_{\p\Sigma} \frac{1}{K} E_\perp \; . \label{eq:Hedge} \ee
As in the Maxwell case, one can write $K$ in terms of an integral kernel constructed from the Green's function for the $S$-Laplacian.

\subsection{Horizon limit}
\label{sec:horlim}

In general $K$ is not a local differential operator on $\p\Sigma$, but in the limit where $\p M$ approaches a bifurcate horizon there is a remarkable simplification, with $K$ becoming proportional to the Laplacian on $\p\Sigma$. The setup for this section is identical to that of section 2.5 in \cite{Ball:2024hqe}. We embed our static Lorentzian manifold $M$ in a spacetime with a bifurcate Killing horizon with respect to $\p_t$, and we take our boundary $\p M$ to lie a proper distance $\eps$ from the horizon (in the direction orthogonal to $\p_t$). In each time slice we set up Gaussian normal coordinates in a neighborhood of the bifurcation surface, so that the full static metric is
\be \ba ds^2 & = g_{tt} dt^2 + g_{ij} dx^i dx^j  = g_{tt} dt^2 + dr^2 + g_{ab} dx^a dx^b \; . \ea \label{eq:staticmetrichorizon}\ee
All components are independent of $t$. The boundary $\p M$ is at $r = \eps$, and the bifurcation surface is at $r=0$. The presence of the static bifurcation surface implies \cite{Solodukhin:2011gn}
\be g_{tt} = -\kappa^2 r^2 + \mc{O}(r^4) \ee
where $\kappa$ is the surface gravity with respect to $\p_t$. The zeroth law of black hole mechanics \cite{Bardeen:1973gs} states that the surface gravity is constant on a stationary horizon, so $g_{tt}$ is independent of the transverse coordinates $x^a$ at leading order in $r$. For the factor $S$ this means
\be S = \sqrt{-g^{tt}} = \frac{1}{\kappa r} + \mc{O}(r) \; . \ee
Another useful fact is that the bifurcation surface of a stationary horizon has vanishing extrinsic curvature \cite{Solodukhin:2011gn}, i.e. the normal derivative of the transverse metric vanishes,
\be 0 = \mc{L}_{\p_r} g_{ab}|_{r=0} = \p_r g_{ab}|_{r=0} \; . \ee
This implies
\be g_{ab} = g_{ab}|_{r=0} + \mc{O}(r^2) \; . \ee
We will work at leading order in $r$, allowing us to drop terms like $S^{-1} \p_a S$ and $\p_r g_{ab}$. We are interested in solving for $\beta$ on $\Sigma$ given data on $\p\Sigma$. Recall the equations are
\be \label{eq:beta-eqns} d^\ddagger \beta = 0 \qquad {\rm and} \qquad d^\ddagger d\beta = 0 \; , \ee
where $d$ is the exterior derivative on $\Sigma$ and $d^\ddagger = \frac{1}{S} d^\dag S$ is defined in \eqref{eq:twistdiff}. In terms of explicit (upper) components these equations are respectively
\be \label{eq:S-c-c-comp} 0 = -\frac{1}{S \sqrt{g_{ij}}} \p_j \left( S\sqrt{g_{ij}} \, \beta^{ji_1\dots i_{p-2}} \right) \ee
and
\be 0 =- \frac{p}{S \sqrt{g_{ij}}} \p_k \left( g^{k\ell} g^{i_1j_1} \dots g^{i_{p-1}j_{p-1}}S\sqrt{g_{ij}} \, \p_{[\ell} \beta_{j_1\dots j_{p-1}]} \right) \; . \ee
The awkward inclusion of many upper metrics in the second equation is the cost of avoiding Christoffel symbols. We relegate most of the calculation to appendix \ref{app:horiz-lim}. There we expand in eigenmodes of the Laplacian $\Delta_{p-1}^{\p\Sigma}$ for $(p-1)$-forms on $\p\Sigma$ and we show that the mode coefficients obey a simple equation at leading order in small $r$. Explicitly, we write
\be \beta_{a_1\dots a_{p-1}}(r, x^a) = \sum_n \beta_n(r) \omega_{n,a_1\dots a_{p-1}}(x^a) \; . \ee
with
\be \int_{\p\Sigma} \omega_m \wedge *_{\p\Sigma} \omega_n = \delta_{mn} \; , \ee
and
\be \Delta_{p-1}^{\p\Sigma} \omega_n = \lam_n \omega_n \ee
with non-negative eigenvalues $\lam_n$. The mode coefficients satisfy
\be 0 \approx -\frac{1}{S} \p_r \left(S \p_r \beta_n\right) + \lam_n \beta_n \; , \ee
where $\approx$ indicates equality at leading order in $r$. Finally plugging in $S = \frac{1}{\kappa r} + \mc{O}(r)$ gives
\be 0 \approx -r \p_r \left(\frac{1}{r} \p_r \beta_n\right) + \lam_n \beta_n \; . \ee
This is identical to (2.69) from \cite{Ball:2024hqe} for the $p=1$ case. The asymptotic form of the solution at small $r$ is
\be \beta_n(r) \propto 1 + \half \lam_n r^2 \log \sqrt{\lam_n} r + C r^2 + \mc{O}(r^3) \; . \ee
The coefficient $C$ is determined by regularity in the bulk, and then the only remaining freedom is in the overall normalization of the solution. The operator $K$, which by definition maps $i_{\p\Sigma} \beta$ to $E_\perp$, then acts as
\be \frac{E_{\perp,n}}{\beta_n|_{r=\eps}} = \frac{-\sqrt{-g^{tt}} \p_r \beta_n}{\beta_n} \Big|_{r=\eps} = \frac{\lam_n}{\kappa} \log \frac{1}{\eps} + \mc{O}(\eps^0) \; . \ee
From this we conclude in the $\eps\to 0$ limit
\be K \quad \longleftrightarrow \quad \frac{\log\eps^{-1}}{\kappa} \Delta^{\p\Sigma}_{p-1} \; . \ee

\section{Quantum Edge Modes}
\label{sec:quantum}

In this section we quantize our edge theory, compute its partition function, argue for its shrinkability, and show that our results confirm some conjectures and resolve some mysteries in the literature.

\subsection{Edge Partition Function}
\label{sec:EdgeP}

In this subsection we compute the thermal partition function in the special case where $\Sigma$ has the topology of a $(D-1)$-ball. Most of the topological factors then drop out. The most important example in this class of manifolds is the de Sitter static patch. We work in the horizon limit of section \ref{sec:horlim}, so throughout this section we set
\be K = \frac{\log\eps^{-1}}{\kappa} \Delta_{p-1}^{\p\Sigma} \; . \ee
For uniformity of discussion we also assume $2 \le p \le D-2$, although most formulas apply to $p=0,1$ as well without modification. Upon quantization, the factorization of phase space implies a tensor factorization of Hilbert space,
\be \mc{H}_{\rm DEM} = \mc{H}_{\rm bulk} \otimes \mc{H}_{\rm edge} \; . \ee
Since the Hamiltonian splits as well we have\footnote{The thermal $\beta$ here is unrelated to the edge mode field $\beta$. We hope this does not cause any confusion.}
\be \ba Z_{\rm DEM}(\beta) & = \Tr \, e^{-\beta H} \\
& = \Tr_{\rm bulk} e^{-\beta H_{\rm bulk}} \Tr_{\rm edge} e^{-\beta H_{\rm edge}} \\
& = Z_{\rm bulk}(\beta) Z_{\rm edge}(\beta) \; . \ea \label{eq:ZDEM}\ee
As discussed above, the bulk degrees of freedom are precisely those of the PMC boundary condition, so we can identify $Z_{\rm bulk}(\beta) = Z_{\rm PMC}(\beta)$. Our interest is in $Z_{\rm edge}(\beta)$. We will evaluate it as a canonical trace in the $\ket{E_\perp}$ basis of edge states with definite $E_\perp$. There are no harmonic $(p-1)$-forms on $\p\Sigma = S^{D-2}$ for $2 \le p \le D-2$ so $K$ has no zero modes and $E_\perp$ can be any co-closed form on $\p\Sigma$. Apparently also $i_{\p\Sigma} \beta = K^{-1} E_\perp$ can be any co-closed form. The same argument implies that $i_{\p\Sigma} \alpha$ can be any co-closed form. It will be convenient to expand in orthonormal eigenmodes of the Laplacian on $\p\Sigma$,
\be \alpha(x^a) = \sum_n \alpha_n \omega_n(x^a) \; , \qquad \beta(x^a) = \sum_n \beta_n \omega_n(x^a) \; , \qquad E_\perp(x^a) = \sum_n E_{\perp,n} \omega_n(x^a) \; . \ee
Here $\omega_n(x^a)$ is a co-closed $(p-1)$-form on $\p\Sigma$ satisfying
\be \Delta_{p-1}^{\p\Sigma} \omega_n(x^a) = \lam_n \omega_n(x^a) \ee
with $\lam_n$ the $n$th Laplacian eigenvalue and
\be \int_{\p\Sigma} \omega_m \wedge *_{\p\Sigma} \omega_n = \delta_{mn} \; . \ee
Plugging into the edge symplectic form gives
\be \Omega_{\rm edge} = \sum_n \delta\alpha_n \delta E_{\perp,n} \; , \ee
while the edge Hamiltonian reads
\be H_{\rm edge} = \sum_n \frac{\kappa E_{\perp,n}^2}{2\lam_n\log\eps^{-1}} \; . \ee
We see that $\alpha_n$ and $E_{\perp,n}$ are conjugate variables, with each mode amounting to a copy of free 1D quantum mechanics. We define a state $\ket{E_{\perp,n}}$ for each individual mode, normalized as
\be \inner{E_{\perp,m}}{E'_{\perp,n}} = 2\pi \delta_{mn} \, \delta(E_{\perp,m} - E'_{\perp,n}) \; . \ee
Note that we can understand
\be \inner{E_{\perp,n}}{E_{\perp,n}} = 2\pi \delta(0) = \int_{-\infty}^\infty d\alpha_n \ee
as the size of the range of $\alpha_n$. The full state is
\be \ket{E_\perp} = \prod_n \ket{E_{\perp,n}} \; . \ee
With this normalization, the resolution of the identity is
\be \int \mc{D}E_\perp \ket{E_\perp}\bra{E_\perp} = \int \prod_n \frac{dE_{\perp,n}}{2\pi} \ket{E_{\perp,n}}\bra{E_{\perp,n}}. \ee
Applying all this to the partition function, we have
\be \ba Z_{\rm edge}(\beta) & = \int \mc{D}E_\perp \bra{E_\perp} e^{-\beta H_{\rm edge}} \ket{E_\perp} \\
& = \int \prod_n \frac{dE_{\perp,n}}{2\pi\mu} \mu \inner{E_{\perp,n}}{E_{\perp,n}} \exp\left( \frac{-\kappa\beta E_{\perp,n}^2}{2\lam_n\log\eps^{-1}} \right) \\
& = \left( \prod_n \frac{\mu}{\sqrt{2\pi}} \int_{-\infty}^\infty d\alpha_n \right) \left( \prod_n \frac{\lam_n\log\eps^{-1}}{\mu^2\kappa\beta} \right)^{\frac12} \\
& = \big| \mc{C}^{p-1}(\p\Sigma) \big| \, \det\left(\frac{\log\eps^{-1}}{\mu^2\kappa\beta} \Delta_{p-1}^{T,\p\Sigma}\right)^{\frac12}. \ea \ee
The expressions in the last line are defined by the penultimate line. We have introduced an arbitrary factor $\mu$ with dimensions of mass so that the infinite products are separately dimensionless. It essentially serves as a renormalization scale. In the final determinant we write $\Delta_{p-1}^{T,\p\Sigma}$ for the restriction of $\Delta_{p-1}^{\p\Sigma}$ to the co-closed, or ``transverse", subspace. The space of $i_{\p\Sigma}\alpha$, i.e. the space of co-closed forms $\mc{C}^{p-1}(\p\Sigma)$, clearly has infinite volume. Thus $Z_{\rm edge}(\beta)$ is ill-defined. Just as in Maxwell \cite{Ball:2024hqe}, our true finite object of interest is actually a quotient of $Z_{\rm edge}(\beta)$, namely
\be \bar Z_{\rm edge}(\beta) \equiv \frac{1}{|\mc{G}^{\p\Sigma}|} Z_{\rm edge}(\beta) \; . \ee
Here $\mc{G}^{\p\Sigma}$ is the group of large gauge transformations with a particular measure, motivated in part by locality and by analogy with \eqref{eq:fullG} for the full path integral measure on a closed manifold. Explicitly, we define
\be \label{eq:fullGlarge} |\mc{G}^{\p\Sigma}| \equiv |\mc{G}^{\p\Sigma}_{p-1} / (\mc{G}^{\p\Sigma}_{p-2} / \dots)| = |\mc{G}^{\p\Sigma}_{\rm nontriv}| \, |U(1)|^{(-)^{p+1}} \prod_{k=0}^{p-2} \left( \det\hspace{0mm}'(\mu^{-2} \Delta_k^{T,\p\Sigma})^{-\frac12} \right)^{(-)^{p-k+1}} \; , \ee
where $\mc{G}^{\p\Sigma}_k$ for $k>0$ acts by a shift $d\lam$ where $\lam \in \Omega^k(\p\Sigma)$ is \textit{any} one-form on $\p\Sigma$, and $\mc{G}^{\p\Sigma}_0$ acts by a shift $d\phi$ where $\phi$ is any $U(1)$-valued function on $\p\Sigma$. We write $\mc{G}^{\p\Sigma}_{\rm nontriv}$ for the part of $\mc{G}^{\p\Sigma}_{p-1}$ with $\lam\in\mc{C}^{p-1}(\p\Sigma)$, meaning that $d\lam\ne 0$ (unless $\lam$ itself vanishes). The appearance of $\mu$ in the determinants follows from using the dimensionless measure $\mc{D}\lam = \prod_n \frac{\mu^{p-k}}{\sqrt{2\pi}} d\lam_n$ for the $k$-forms in $\mc{G}^{\p\Sigma}_k$. The $k=0$ case also induces a measure on $U(1)$. We know $\phi \sim \phi + \frac{2\pi}{q}$, and the zero mode part of $\phi$ is $\phi_0 V_{\p\Sigma}^{-1/2}$, so apparently $\phi_0 \sim \phi_0 + \frac{2\pi}{q} V_{\p\Sigma}^{1/2}$. Integrating freely over the zero mode coefficient then gives
\be |U(1)| = \int_0^{\frac{2\pi}{q} \sqrt{V_{\p\Sigma}}} \frac{\mu^p}{\sqrt{2\pi}} d\phi_0 = \frac{\mu^p}{q} \sqrt{2\pi V_{\p\Sigma}} \; . \ee
Overall we have
\be \label{eq:Zbedge-penul} \ba \bar Z_{\rm edge}(\beta) & = \frac{|\mc{C}^{p-1}(\p\Sigma)|}{|\mc{G}^{\p\Sigma}|} \, \det\left(\frac{\log\eps^{-1}}{\mu^2\kappa\beta} \Delta_{p-1}^{T,\p\Sigma}\right)^{\frac12} \\
& = \left( |U(1)|^{(-)^p} \prod_{k=0}^{p-2} \left( \det\hspace{0mm}'(\mu^{-2} \Delta_k^{T,\p\Sigma})^{-\frac12} \right)^{(-)^{p-k}} \right) \det\left( \frac{\log\eps^{-1}}{\mu^2\kappa\beta} \Delta_{p-1}^{T,\p\Sigma} \right)^{\frac12} \\
& = \left( \frac{\log\eps^{-1}}{\kappa\beta} \right)^{\half (-)^p + \rm{anomaly}} \det(\mu^{-2} \Delta_{p-1}^{T,\p\Sigma})^{\frac12} |U(1)|^{(-)^p} \prod_{k=0}^{p-2} \left( \det\hspace{0mm}'(\mu^{-2} \Delta_k^{T,\p\Sigma})^{-\frac12} \right)^{(-)^{p-k}} \ea \ee
where the factor $\left( \frac{\log\eps^{-1}}{\kappa\beta} \right)^{\half(-)^p + \rm{anomaly}}$ comes from scaling the factor $\frac{\log\eps^{-1}}{\kappa\beta}$ out of the $(p-1)$-form determinant. The general rule for pulling out a constant from a (full) $p$-form Laplacian $\Delta_p$ is \cite{Rosenberg}
\be \det\hspace{0mm}'(a \Delta_p) = a^{-\dim\ker\Delta_p + \rm{anomaly}} \det\hspace{0mm}'(\Delta_p) \; , \ee
where the anomaly part is present only in even dimensions. The dimension of the kernel of the Laplacian on a closed manifold is also known as the Betti number $b_p$. The rule for a transverse Laplacian can be deduced from the recursive relation \eqref{eq:detid}. One finds
\be \det\hspace{0mm}'(a \Delta^T_p) = a^{-\sum_{k=0}^p (-)^{p-k} b_k + \rm{anomaly}} \det\hspace{0mm}'(\Delta^T_p) \; . \ee
In the present case where $\p\Sigma$ is topologically a sphere, the only nonzero Betti numbers are $b_0 = b_{D-2} = 1$. Aside from this factor, we recognize the rest of the final expression in \eqref{eq:Zbedge-penul} as the reciprocal of the partition function $\mc{Z}_{\rm PI}^{p-1}[\p\Sigma]$ of a $(p-1)$-form gauge theory on $\p\Sigma$ with fundamental charge $q$,\footnote{In general the fundamental charge of $p$-form gauge theory in $D$ dimensions has mass dimension $[q] = -\half D + p + 1$, so it is the same mass dimension as for $(p-1)$-form gauge theory in $D-2$ dimensions.} so we can finally write
\be \label{eq:Zbedge-final} \bar Z_{\rm edge}(\beta) = \frac{\left( \frac{\log\eps^{-1}}{\kappa\beta} \right)^{\half(-)^p + \rm{anomaly}}}{\mc{Z}_{\rm PI}^{p-1}[\p\Sigma]}\; . \ee
This is one of our main results. Henceforth we always normalize time such that $\kappa=1$, in which case a non-singular Euclidean manifold corresponds to $\beta=2\pi$. For completeness we define
\be \bar Z_{\rm DEM}(\beta) \equiv Z_{\rm bulk}(\beta) \bar Z_{\rm edge}(\beta)\; . \ee
The quotient by $\mc{G}^{\p\Sigma}$ in the definition of $\bar Z_{\rm DEM}(\beta)$ is natural from the perspectives of locality and gauge invariance (see e.g. section 4.3 of \cite{Ball:2024hqe}), but ultimately our interest in it derives from the fact that it seems to define a shrinkable boundary condition, as we discuss next.

\subsection{Shrinkability of the DEM boundary condition}
\label{sec:shrink}

In \eqref{eq:ZDEM} we opted to compute the thermal partition function as a canonical trace. Standard arguments imply that it can be alternatively computed by the path integral on the Euclidean manifold obtained from the Lorentzian manifold $M$ \eqref{eq:staticmetrichorizon} by a Wick rotation $t\to -i \tau$ together with the identification $\tau \sim \tau + 2\pi$ (recall that we are setting $\kappa=1$). Under this Wick rotation, the timelike boundary a proper distance $\eps$ from the horizon is mapped to a surface $S^1\times\p\Sigma$ (where $S^1$ has radius $\eps$) excising the origin. Throughout this section we will use $\mc{M}_\eps$ to denote this Euclidean manifold, and $\mc{M}$ to denote the corresponding smooth Euclidean manifold with no hole cut out.

\begin{figure}[H]
    \centering
    \includegraphics[width=0.6\textwidth]{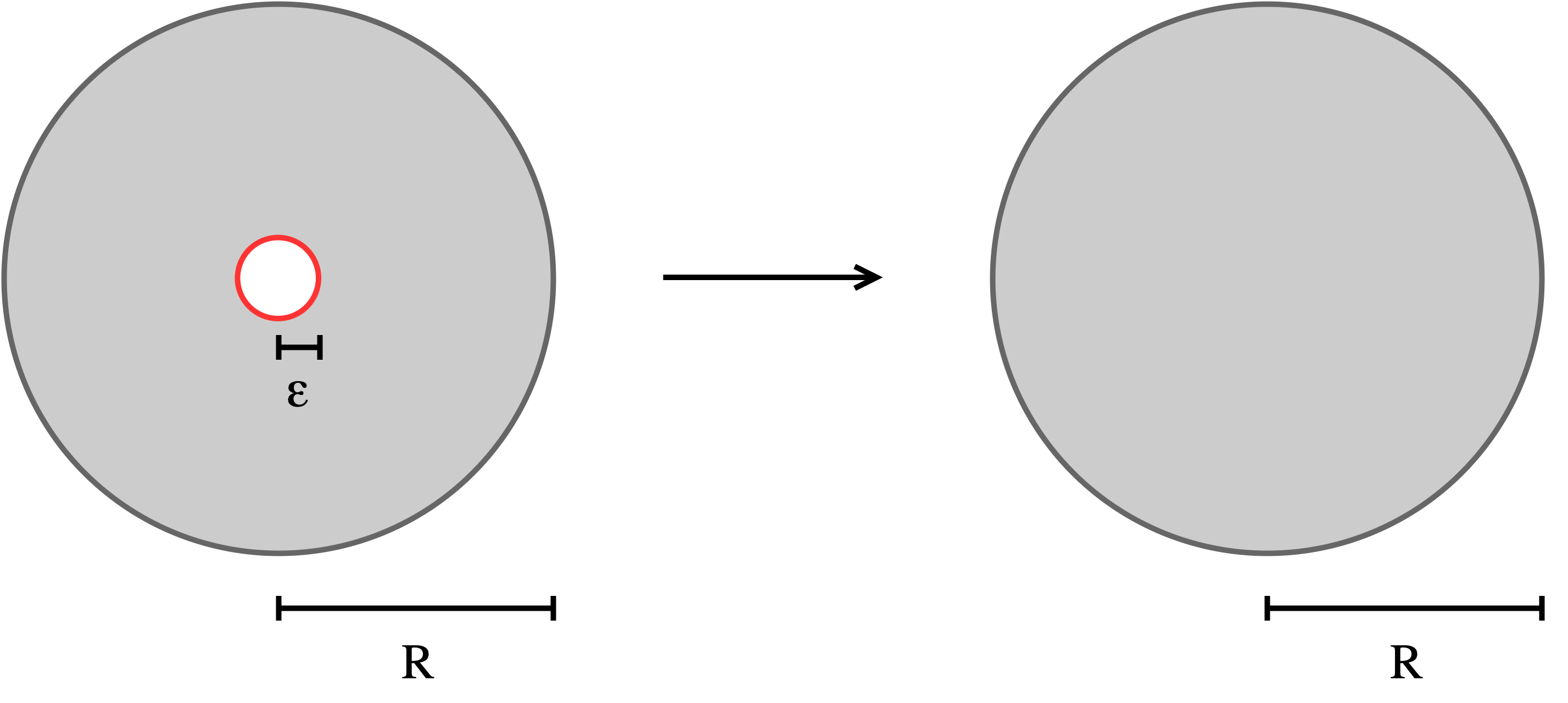}
        \caption{With an appropriate boundary condition, the annulus with inner radius $\eps$ can recover the disk as $\eps\to 0$.}
    \label{fig:ann-shrink}
\end{figure}

We call a boundary condition {\it shrinkable} if when applied to an infinitesimally small hole the result is as if there were no hole at all. Said differently, in general one expects an infinitesimally small hole with a given boundary condition to be effectively described by some defect operator on the corresponding manifold with no hole,\footnote{This may require local counterterms on the defect geometry itself, in addition to the usual counterterms on the manifold and its boundary.} and the boundary condition is shrinkable if the defect operator is the identity. Figure \ref{fig:ann-shrink} shows the 2D example of recovering the disk from the annulus in this way. We will be interested in the shrinkability of our DEM boundary condition for the manifold $\mc{M}_\eps$ as $\eps\to 0$. We refer the reader to \cite{Ball:2024hqe} for a more detailed discussion of shrinkability in the context of edge modes, and to \cite{Jafferis:2019wkd, Hung:2019bnq, Agia:2022srj, Agia:2024wxx} for other discussions.

\subsubsection{Comparing with $S^D$ with no hole}\label{sec:ZPISDnohole}

We now demonstrate that the quantum boundary condition \eqref{eq:DEMBC} defining $\bar Z_{\rm DEM}(\beta)$ is shrinkable, at least at the level of the partition function, in the case where $\mc{M}$ is the round sphere $S^D$ of radius $R$. The relevant Lorentzian geometry is a static patch of de Sitter space in $D$ dimensions ($dS_D$), described by the metric
\begin{align}\label{eq:staticpatchmetric}
    ds^2 = - \left(R^2-\rho^2\right)dt^2 + \frac{d\rho^2}{1-\frac{\rho^2}{R^2} } +\rho^2 d\Omega_{D-2}^2 \;, \qquad 0\leq \rho <R \;. 
\end{align}
By performing a Wick rotation $t \to -i\tau$ and imposing the periodic identification $\tau \sim \tau + 2\pi$, this metric becomes that of a round sphere $S^D$ of radius $R$, with the horizon at $\rho = R$ mapped to the Euclidean origin. To apply our analysis in section \ref{sec:CovPhaseEdge} to this case, we excise the horizon at a proper distance $\eps$, resulting in a static spacetime with a timelike boundary.

The partition function of $U(1)$ $p$-form gauge theory on the round sphere $S^D$ of radius $R$ with fundamental charge $q$ takes the form \cite{Cappelli:2000fe}
\begin{align}\label{eq:masslessPIdet}
	\mc{Z}^{p}_{\rm PI} \left[S^D\right]
	= & |U(1)|^{(-)^p}  \left( \det\hspace{0mm}'\frac{\Delta_{0}}{\mu^2}\right) ^{-\frac{(-)^{p}}{2}}  \prod_{k=1}^{p}\left( \det\frac{\Delta^T_{k}}{\mu^2}\right) ^{-\frac{(-)^{p-k}}{2}} \; .
\end{align}
The prime in $\det'\frac{\Delta_0}{\mu^2}$ denotes the exclusion of the zero mode from the determinant. The group volume factor $|U(1)|$ takes the form \eqref{eq:U(1)vol} with $V_\mc{M}= R^D{\rm Vol}\left(S^D\right)$ where ${\rm Vol}\left(S^D\right)=\frac{2\pi^{\frac{D+1}{2}}}{\Gamma\left(\frac{D+1}{2} \right)}$ is the volume of a unit round $S^D$.\footnote{We note that \cite{Cappelli:2000fe} did not take into account correctly the volume of $U(1)$ and the fundamental charge $q$ in their zero mode integration.} In this and subsequent sections, we use $\mc{Z}$ (as opposed to $Z$) to emphasize quantities defined without a brick wall.

In \cite{David:2021wrw,Mukherjee:2023ihb}, it was shown that \eqref{eq:masslessPIdet} naturally splits into a ``bulk" and an ``edge" part
\begin{align}\label{eq:masslessPIbulkedge}
    \mc{Z}^{p}_{\rm PI} \left[S^D\right] = \mc{Z}^p_{\rm bulk}\left[dS_D\right]\mc{Z}^p_{\rm edge}\left[S^D\right] \; .
\end{align}
The bulk part is given by the (unregularized) integral
\begin{align}\label{eq:Zbulk}
    \log \mc{Z}^p_{\rm bulk}\left[dS_D\right]= \int_0^\infty \frac{dt}{2t}\frac{1+e^{-\frac{t}{\mu R}}}{1-e^{-\frac{t}{\mu R}}} \chi_p \left( \frac{t}{\mu R}\right) \; , 
\end{align}
where the quantity 
\begin{align}\label{eq:HCcharpgauge}
    \chi_p(t) = - (-)^p+\sum_{k=0}^p (-)^{k+p} \binom{D-1}{k}   \frac{e^{-k t} +e^{-\left(D-1-k\right)t}}{\left(1-e^{-t} \right)^{D-1}} 
\end{align}
is the Harish-Chandra character for a massless $p$-form gauge field in $dS_D$. As pointed out in \cite{Anninos:2020hfj} and recalled in appendix \ref{sec:Zbulk}, \eqref{eq:Zbulk} or more generally \eqref{appeq:idealgas} can be independently defined in a $dS_D$ static patch at any inverse temperature $\beta $, with a Lorentzian interpretation as a ``quasicanonical" trace over local propagating degrees of freedom in the static patch. The formula \eqref{eq:Zbulk} corresponds to $\beta=2\pi$. In \cite{David:2021wrw}, the physical interpretation of the edge part $\mc{Z}^p_{\rm edge}\left[S^D\right]$ was not as well understood, and $\mc{Z}^p_{\rm edge}\left[S^D\right]$ was presented in terms of character integrals of the form \eqref{eq:Zbulk} on a series of lower-dimensional de Sitter spaces, namely $dS_{D-2}, dS_{D-4}, \cdots , dS_{D-2p}$ (when $2p<D$). Later in \cite{Mukherjee:2023ihb}, (the non-zero mode part of) $\mc{Z}^p_{\rm edge}\left[S^D\right]$ was recognized to be the reciprocal of the partition function of a massless $(p-1)$-form gauge theory on $S^{D-2}$. In appendix \ref{app:masslesspformSD}, we re-derive the split \eqref{eq:masslessPIbulkedge} following the approach in \cite{Anninos:2020hfj}. We show that including the dependence on the fundamental charge $q$ in \eqref{eq:U(1)vol}, we have exactly
\be \mc{Z}^p_{\rm edge}\left[S^D\right] = \frac{1}{\mc{Z}^{p-1}_{\rm PI} \left[S^{D-2}\right]} \; . \label{eq:ZedgeSd-2} \ee

\paragraph{Shrinking the hole}

We compare this with our $\bar Z_{\rm DEM}(\beta)$ for the brick-wall-regulated de Sitter static patch, at temperature $\beta=2\pi$. It reads
\be \bar Z_{\rm DEM}(2\pi) = Z_{\rm bulk}(2\pi) \bar Z_{\rm edge}(2\pi) \;.\label{eq:ZDEMsplitSD} \ee 
We recall that $Z_{\rm bulk}(2\pi)=Z_{\rm PMC}(2\pi)$, i.e. it captures the bulk propagating degrees of freedom satisfying the PMC boundary condition; this is thus expected to agree with $\mc{Z}^p_{\rm bulk}\left[dS_D\right]$ in the $\eps\to 0$ limit (up to possible non-universal terms). This expectation bears out in even dimensions, but in odd dimensions there is a slight discrepancy.\footnote{For Maxwell in certain simple geometries $\log Z_{\rm PMC}$ can be recast as $D-2$ Neumann scalar partition functions, up to a term $\log\frac{Z_D}{Z_N}$, where $Z_{D(N)}$ is the scalar partition function with the Dirichlet (Neumann) boundary condition \cite{Donnelly:2015hxa, Blommaert:2018rsf}. This term is of order $\log\log\eps^{-1}$, which in even $D$ will not affect the coefficient of $\log\mu R$.} On the other hand, we have
\be \bar Z_{\rm edge}(2\pi) = \frac{\left( \frac{\log\eps^{-1}}{2\pi} \right)^{\half(-)^p + \rm{anomaly}}}{\mc{Z}^{p-1}_{\rm PI} \left[S^{D-2}\right]}  = \left( \frac{\log\eps^{-1}}{2\pi} \right)^{\half(-)^p + \rm{anomaly}}\mc{Z}^p_{\rm edge}\left[S^D\right] \; . \ee
In even dimensions the contribution of $\frac{\log\eps^{-1}}{2\pi}$ to $\log\bar Z_{\rm edge}(2\pi)$ is of order $\log\log\eps^{-1}$, which will not affect the coefficient of $\log\mu R$, which is the universal part. In odd dimensions the match is more subtle, but it seems likely that the $\frac{\log\eps^{-1}}{2\pi}$ factor discrepancy will be compensated by another discrepancy between $Z_{\rm bulk}(2\pi)$ and $\mc{Z}_{\rm bulk}^p[dS_D]$, as this is what happens for Maxwell at least in some simple geometries \cite{Donnelly:2015hxa, Blommaert:2018rsf}. To underscore the nontriviality of this match we present some numerical values in section \ref{sec:SDnumerical}.

In addition to its technical tractability, this $S^D$ example is important because of its relevance to de Sitter space quantum gravity. According to the Gibbons-Hawking proposal \cite{Gibbons:1976ue}, $S^D$ partition functions compute quantum corrections to the de Sitter horizon entropy, which provide precision tests for candidate microscopic models in the spirit of \cite{Banerjee:2011jp,Sen:2012dw,Bhattacharyya:2012ye,Sen:2012kpz,Giombi:2013fka,Giombi:2014iua,Gunaydin:2016amv,Giombi:2016pvg,Skvortsov:2017ldz,Anninos:2020hfj}. As shown in \cite{Anninos:2020hfj}, any spinning 1-loop $S^D$ partition functions admit a bulk-edge split as in \eqref{eq:masslessPIbulkedge}. Clarifying the physical meaning of each part may help identify specific sectors of candidate microscopic models. The discussion above establishes that the edge partition function can be interpreted as a thermal partition function of the edge mode sector, corresponding to large gauge transformations residing on the stretched horizon. This provides a concrete physical picture of the edge sector, which was not fully understood in previous studies.

\subsection{Numerical results on $S^D$ for various $D$}
\label{sec:SDnumerical}

In this section, we present numerical results emphasizing the content of \eqref{eq:masslessPIbulkedge}. To that end, we note that using the eigenvalues and degeneracies \eqref{eq:degeneracy} of the Laplacians acting on transverse $p$-forms on a round $S^D$, we can put the massless $U(1)$ $p$-form partition function \eqref{eq:masslessPIdet} into the form (reviewed in appendix \ref{app:masslesspformSD})
\begin{align}\label{eq:SDmasslesspformPI}
	\log \mc{Z}^{p}_{\rm PI} \left[S^D\right] 
 =& \, (-)^{p}\log \frac{\mu^{p+1}\sqrt{2\pi R^D{\rm Vol}\left(S^D\right)}}{q}+ \int_0^\infty \frac{dt}{2t} F\left(\frac{t}{\mu R} \right) 
\end{align}
where the sum
\begin{align}
    F\left(\frac{t}{\mu R} \right) = & \,  \sum_{k=0}^p (-)^{p-k}\left(e^{-k\frac{t}{\mu R}}+e^{-\left(D-1-k\right)\frac{t}{\mu R}}\right) \sum_{n=1}^\infty d^D_{n,(k)} e^{-n\frac{t}{\mu R}} 
\end{align}
can be computed with the explicit form of $d^D_{n,(k)}$ given in \eqref{eq:degeneracy}. The integral 
\begin{align}\label{eq:logZform}
    \int_0^\infty \frac{dt}{2t} F\left(\frac{t}{\mu R} \right) 
\end{align}
is UV-divergent in the region $t\to 0$. A quick way to extract the universal scheme-independent part of \eqref{eq:logZform} is to construct the zeta function
\begin{align}\label{eq:integralzeta}
     \zeta(z)\equiv \frac{1}{\Gamma(z)} \int_0^\infty \frac{dt}{2t}t^z F\left(\frac{t}{\mu R} \right) \; . 
\end{align}
This integral is convergent for large positive $z$. We then evaluate $\zeta'(0)$ and obtain a regularized version of \eqref{eq:SDmasslesspformPI}
\begin{align}\label{eq:charintphysical}
    \log \mc{Z}^{p}_{\rm PI, reg} \left[S^D\right]   &= \log \mc{Z}^{p}_{\rm PI, fin} \left[S^D\right] + \log \mc{Z}^{p}_{\rm PI,log} \left[S^D\right]   \nn\\
    \log \mc{Z}^{p}_{\rm PI, fin} \left[S^D\right]  & = 
    (-)^{p}\log \frac{\sqrt{2\pi R^{D-2p-2}{\rm Vol}\left(S^D\right)}}{q}+\zeta'(0)|_{\mu R=1} \nn\\
    \log \mc{Z}^{p}_{\rm PI,log} \left[S^D\right]   & =\left( \alpha_D+(-)^{p} \left( p+1\right) \right) \log\mu R \; .
\end{align}
In even $D$, the total log-coefficient $\alpha_{D}+(-)^{p} \left( p+1\right)$ is the scheme-independent quantity, where $\alpha_{D}$ can be read off as the coefficient of $\frac{1}{t}$ in the small $t$ expansion of the integrand of \eqref{eq:logZform}:
\begin{align}\label{eq:readoff}
    \frac{1}{2t} F\left(\frac{t}{\mu R} \right) =   \cdots + \frac{\alpha_{D}}{t}+\cdots  \, . 
\end{align}
The term $(-)^p(p+1)$ comes from the non-local zero mode contribution \eqref{eq:SDmasslesspformPI}. In odd $D$, $\log \mc{Z}^{p}_{\rm PI, fin} \left[S^D\right] $ is unambiguously defined, and we will describe its evaluation in more detail in section \ref{sec:oddDsphere}.

If one uses a different regularization scheme, $\mu  $ will be replaced by some combination involving the UV regulator. For example, if one computes the functional determinant of a Laplace-type operator $\mathfrak{D}$ in the heat kernel regularization \cite{Vassilevich:2003xt}
\begin{align}
     \log \left(\det \mathfrak{D}\right)^{-1/2} = \int_0^\infty \frac{d\tau}{2\tau} \, e^{-\epsilon^2/4\tau} \Tr \, e^{- \mathfrak{D}\tau}
\end{align}
with the specific regulator $e^{-\epsilon^2/4\tau}$ as in \cite{Anninos:2020hfj}, $\mu $ would be replaced by $\frac{2e^{-\gamma}}{\epsilon}$ where $\gamma$ is the Euler constant; additionally, there will be divergent terms involving inverse powers of $\epsilon$, which can be absorbed into local counterterms. We refer the reader to \cite{Anninos:2020hfj} for an elaborate discussion on treatment with various regularization schemes.

\subsubsection{Log-coefficients in even $D$}

Using the prescription described above, one can compute the total log-coefficients of $\log \mc{Z}^p_{\rm PI} \left[ S^D\right]$ for arbitrary $0\leq p\leq D-2$ and even $D$. Here we display the explicit results up to $D=10$:
\begin{table}[H]
\centering
\renewcommand{\arraystretch}{1.2}
\begin{tabular}{|c|c|c|c|c|c|c|c|c|c|}
\hline
\diagbox{$D$}{$p$} & $0$ & $1$ & $2$ & $3$ & $4$ & $5$ & $6$ & $7$ & $8$ \\
\hline
$2$  & $\frac{1}{3}$ & & & & & & & & \\
\hline
$4$  & $\frac{29}{90}$ & $-\frac{31}{45}$ & $\frac{209}{90}$ & & & & & & \\
\hline
$6$  & $\frac{1139}{3780}$ & $-\frac{1271}{1890}$ & $\frac{221}{210}$ & $-\frac{5051}{1890}$ & $\frac{16259}{3780}$ & & & & \\
\hline
$8$  & $\frac{32377}{113400}$ & $-\frac{4021}{6300}$ & $\frac{2603}{2520}$ & $-\frac{8051}{5670}$ & $\frac{7643}{2520}$ & $-\frac{29221}{6300}$ & $\frac{712777}{113400}$ & & \\
\hline
$10$ & $\frac{2046263}{7484400}$ & $-\frac{456569}{748440}$ & $\frac{13228}{13365}$ & $-\frac{5233531}{3742200}$ & $\frac{1339661}{748440}$ & $-\frac{12717931}{3742200}$ & $\frac{66688}{13365}$ & $-\frac{4947209}{748440}$ & $\frac{61921463}{7484400}$ \\
\hline
\end{tabular}
\caption{Coefficients of $\log \mc{Z}^{p}_{\rm PI,log} \left[S^D\right]  $ in $D=2,4,6,8,10$.}
\label{tab:PIlog}
\end{table}
\noindent These numerical values have appeared in \cite{David:2021wrw}, and generalize earlier calculations of sphere free energies in \cite{Elizalde:1996nb,Cappelli:2000fe,Raj:2016zjp}.

Since the bulk partition function \eqref{eq:Zbulk} also takes the form \eqref{eq:logZform}, we can extract its contribution to the total log-coefficient by reading off the coefficient of the $O\left( \frac{1}{t}\right)$ term of the small-$t$ expansion of the integrand \eqref{eq:Zbulk}:
\begin{table}[H]
\centering
\renewcommand{\arraystretch}{1.2}
\begin{tabular}{|c|c|c|c|c|c|c|c|c|c|}
\hline
\diagbox{$D$}{$p$} & $0$ & $1$ & $2$ & $3$ & $4$ & $5$ & $6$ & $7$ & $8$ \\
\hline
$4$ & $\frac{29}{90}$ & $-\frac{16}{45}$ & $\frac{29}{90}$ & & & & & & \\
\hline
$6$ & $\frac{1139}{3780}$ & $-\frac{331}{945}$ & $\frac{229}{630}$ & $-\frac{331}{945}$ & $\frac{1139}{3780}$ & & & & \\
\hline
$8$ & $\frac{32377}{113400}$ & $-\frac{1592}{4725}$ & $\frac{545}{1512}$ & $-\frac{1042}{2835}$ & $\frac{545}{1512}$ & $-\frac{1592}{4725}$ & $\frac{32377}{113400}$ & & \\
\hline
$10$ & $\frac{2046263}{7484400}$ & $-\frac{303601}{935550}$ & $\frac{657683}{1871100}$ & $-\frac{342019}{935550}$ & $\frac{276929}{748440}$ & $-\frac{342019}{935550}$ & $\frac{657683}{1871100}$ & $-\frac{303601}{935550}$ & $\frac{2046263}{7484400}$ \\
\hline
\end{tabular}
\caption{Coefficients of $\log \mc{Z}^{p}_{\rm bulk, log} \left[dS_D\right]  $ in $D=4,6,8,10$.}
\label{tab:bulklog}
\end{table}
\noindent In table \ref{tab:bulklog}, it is clear that $\alpha^{\rm bulk}_D$ is invariant under sending $p\to D-2-p$, which can be traced back to the fact that the Harish-Chandra character \eqref{eq:HCcharpgauge} is invariant under this duality map:
\begin{align}\label{eq:HCduality}
    \chi_p(t) = \chi_{D-2-p}(t) \; . 
\end{align}
Finally, since $\mc{Z}^p_{\rm edge}\left[S^D\right]=1/\mc{Z}^{p-1}_{\rm PI} \left[S^{D-2}\right]$, we can compute its log-coefficient again using the same prescription \eqref{eq:charintphysical}, resulting in:
\begin{table}[H]
\centering
\renewcommand{\arraystretch}{1.2}
\begin{tabular}{|c|c|c|c|c|c|c|c|c|c|}
\hline
\diagbox{$D$}{$p$} & $0$ & $1$ & $2$ & $3$ & $4$ & $5$ & $6$ & $7$ &$8$\\
\hline
$4$ & $0$ & $-\frac{1}{3}$ & $2$ & & & & & &\\
\hline
$6$ & $0$ & $-\frac{29}{90}$ & $\frac{31}{45}$ & $-\frac{209}{90}$ & $4$ & & & &\\
\hline
$8$ & $0$ & $-\frac{1139}{3780}$ & $\frac{1271}{1890}$ & $-\frac{221}{210}$ & $\frac{5051}{1890}$ & $-\frac{16259}{3780}$ & $6$ & &\\
\hline
$10$ & $0$ & $-\frac{32377}{113400}$ & $\frac{4021}{6300}$ & $-\frac{2603}{2520}$ & $\frac{8051}{5670}$ & $-\frac{7643}{2520}$ & $\frac{29221}{6300}$ & $-\frac{712777}{113400}$ & $8$ \\
\hline
\end{tabular}
\caption{Coefficients of $\log \mc{Z}^{p}_{\rm edge, log} \left[S^D\right]  $ in $D=4,6,8,10$.}
\label{tab:edgelog}
\end{table}
\noindent Except for $p=0$ and $p=D-2$, table \ref{tab:edgelog} is the same as table \ref{tab:PIlog} after shifting $p\to p-1$ and $D\to D-2$ and taking the negatives. One can readily check that $\log\mc{Z}^{p}_{\rm PI, log} \left[S^D\right] =\log\mc{Z}^{p}_{\rm bulk, log} \left[dS_D\right]  +\log\mc{Z}^{p}_{\rm edge, log} \left[S^D\right]  $.

\paragraph{Comments on duality anomaly}

In tables \ref{tab:PIlog} and \ref{tab:edgelog}, the coefficients for the $p$-form differ from those of the dual $(D-2-p)$-form by $(-)^p \left( D-2-2p\right)$. This ``duality anomaly" in even $D$ was first observed in \cite{Duff:1980qv}, where it was found that the trace anomalies for the 0- and 2-form in $D=4$ differ by the Euler number $\chi$ of the manifold. Later this was proved more generally in \cite{Bastianelli:2005vk,Donnelly:2016mlc}.

For the case of $S^D$, this has been obtained explicitly in \cite{David:2021wrw} and \cite{Raj:2016zjp}. In fact, one can see this more directly as follows. For a $p$-form on $S^{2N+2}$ with $p\geq N+1$, since the eigenvalues and degeneracies \eqref{eq:degeneracy} of the transverse Laplacians are invariant upon sending $p\to D-1-p$, we can rewrite \eqref{eq:masslessPIdet} as follows:
\begin{align}\label{eq:PIlargep}
	&\mc{Z}^{p}_{\rm PI} \left[S^D\right] \nn\\
	= & \left( \frac{\mu^{p+1}\sqrt{2\pi R^D{\rm Vol}\left(S^D\right)}}{q}\right)^{(-)^p}   \left( \prod_{k=0}^{N}\left( \det\frac{\Delta^T_{k}}{\mu^2}\right) ^{-\frac{(-)^{p-k}}{2}}  \right) \left( \prod_{k=N+1}^{p}\left( \det\frac{\Delta^T_{D-1-k}}{\mu^2}\right) ^{-\frac{(-)^{p-k}}{2}} \right)  \nn\\
	= & \left( \frac{\mu^{p+1}\sqrt{2\pi R^D{\rm Vol}\left(S^D\right)}}{q}\right)^{(-)^p}   \left( \prod_{k=0}^{N}\left( \det\frac{\Delta^T_{k}}{\mu^2}\right) ^{-\frac{(-)^{p-k}}{2}}  \right) \left( \prod_{k=D-1-p}^{N}\left( \det\frac{\Delta^T_{k}}{\mu^2}\right) ^{-\frac{(-)^{p-D+1+ k}}{2}} \right) \nn\\
	= & \left( \frac{\mu^{\tilde p}\sqrt{2\pi R^D{\rm Vol}\left(S^D\right)}}{q \mu^{D-2-2p}}\right)^{(-)^{\tilde p}} \left( \det\hspace{0mm}'\frac{\Delta_{0}}{\mu^2}\right) ^{-\frac{(-)^{\tilde p}}{2}} \prod_{k=1}^{\tilde p}\left( \det\frac{\Delta^T_{k}}{\mu^2}\right) ^{-\frac{(-)^{\tilde p-k}}{2}} \;. 
\end{align}
We suppressed in the first two equalities the prime on the 0-form determinant, which is restored in the last line. In the final expression, we have written in terms of $\tilde p=D-2- p$ (so that $\tilde p < N+1$). Now, consider the partition function for a dual $\tilde p$-form with charge $\tilde q \equiv 2\pi/q$:
\begin{align}\label{eq:dualPI}
	\mc{Z}^{\tilde p}_{\rm PI} \left[S^D\right] = \left( \frac{\mu^{\tilde p}\sqrt{2\pi R^D{\rm Vol}\left(S^D\right)}}{\tilde q }\right)^{(-)^{\tilde p}} \left( \det\hspace{0mm}'\frac{\Delta_{0}}{\mu^2}\right) ^{-\frac{(-)^{\tilde p}}{2}} \prod_{k=1}^{\tilde p}\left( \det\frac{\Delta^T_{k}}{\mu^2}\right) ^{-\frac{(-)^{\tilde p-k}}{2}} \;. 
\end{align}
Taking the ratio between \eqref{eq:PIlargep} and \eqref{eq:dualPI}, we find
\begin{align}
    \log \frac{\mc{Z}^{p}_{\rm PI} \left[S^D\right]}{\mc{Z}^{\tilde p}_{\rm PI} \left[S^D\right]} = &(-)^{\tilde p +1} 2 \log \left(  \mu^{\frac{D-2}{2}-p} \sqrt{\frac{q }{\tilde q}}\right) \nn\\
    =& (-)^{\tilde p +1} (D-2-2p) \log \mu R+ (-)^{\tilde p +1}  \log  \frac{q }{\tilde q R^{D-2-2p}}
 \;. 
\end{align}
The first line is the special case of (1.1) in \cite{Donnelly:2016mlc}, where $M=S^D$ and $\chi(S^D)=2$. The first term on the second line explains the difference in log-coefficients for dual pairs. What is new and interesting here is that the anomaly is completely captured by the edge partition function, as encapsulated in table \ref{tab:edgelog}.

\subsubsection{UV-finite parts in odd $D$}\label{sec:oddDsphere}

In odd $D$, the total log-divergent term $\log \mc{Z}^{p}_{\rm PI,log} \left[S^D\right]  $ in \eqref{eq:charintphysical} vanishes,\footnote{This is related to the fact that in odd $D$, there is no integral of curvature invariants of the (schematic) form $\int d^Dx \sqrt{g} \mathcal{R}^k$ that is dimensionless on a compact manifold that can act as a local counterterm.} and the finite part $\log \mc{Z}^{p}_{\rm PI, fin} \left[S^D\right] $ is unambiguously defined. This is also true separately for $\log \mc{Z}^{p}_{\rm bulk} \left[S^D\right]$ and $\log \mc{Z}^{p}_{\rm edge} \left[S^D\right]$.\footnote{For the former, one finds that the $O\left( \frac{1}{t}\right)$ term is absent in the small-$t$ expansion \eqref{eq:readoff} of the integrand \eqref{eq:Zbulk}.} This implies that there cannot be any dependence on $\mu R$ in the final answer, and we can thus set $\mu R=1$ in the following.

As explained in appendix C of \cite{Anninos:2020hfj}, we can directly express zeta functions of the form \eqref{eq:integralzeta} in terms of the Hurwitz zeta function $\zeta_H(z,a) = \sum_{n=0}^\infty (n+a)^{-z}$ as follows. We first expand $F(t)$ in the integrand \eqref{eq:integralzeta} in powers of $e^{-t}$,
\begin{align}
    \zeta(z) = \frac{1}{\Gamma(z)}\int_0^\infty \frac{dt}{2t} t^z \sum_{n=0}^\infty Q(n) \, e^{-(n+\Delta) t} \; .
\end{align}
Together with the integral representation of the Hurwitz zeta function
\begin{align}
    \zeta_H(z,a) = \sum_{n=0}^\infty \int_0^\infty \frac{dt}{t} t^z \, e^{-(n+a)t} \; , 
\end{align}
we can then write
\begin{align}
    \zeta(z) = \frac12 Q(\delta_z - \Delta) \zeta_H(z,\Delta) \; .
\end{align}
Here $\delta_z$ is a shifting operator acting as $\delta_z^n \zeta_H(z,\Delta)=\zeta_H(z-n,\Delta)$. For example, if $Q(n)=n^2$, we have $Q(\delta_z - \Delta) \zeta_H(z,\Delta) = (\delta_z^2 - 2 \Delta \delta_z+\Delta^2)\zeta_H(z,\Delta) = \zeta_H(z-2,\Delta) - 2 \Delta \zeta_H(z-1,\Delta)+\Delta^2 \zeta_H(z,\Delta)$. Noting these, we obtain the results in $D=3,5,7$, summarized in tables \ref{tab:gammaD=3}-\ref{tab:gammaD=7}. 

\begin{table}[H]
\centering
\renewcommand{\arraystretch}{1.5} 
\begin{tabular}{|c|c|c|c|}
\hline
$p$ & $\log \mc{Z}^p_{\rm bulk, fin}$ & $\log \mc{Z}^p_{\rm edge, fin}$ & $\log \mc{Z}^p_{\rm PI, fin}=\log \mc{Z}^p_{\rm bulk, fin}+\log \mc{Z}^p_{\rm edge, fin}$ \\
\hline
$0$ & $-\frac{\zeta (3)}{4 \pi^2}$ & $\log \frac{2 \pi \sqrt{R}}{q}$ & $\log \frac{2 \pi \sqrt{R}}{q} - \frac{\zeta(3)}{4 \pi^2}$ \\
\hline
$1$ & $-\frac{\zeta (3)}{4 \pi^2}$ & $\log  q \sqrt{R}$ & $\log  q \sqrt{R} - \frac{\zeta(3)}{4 \pi^2}$ \\
\hline
\end{tabular}
\caption{$D=3$}
\label{tab:gammaD=3}
\end{table}

\begin{table}[H]
\centering
\renewcommand{\arraystretch}{1.5} 
\begin{tabular}{|c|c|c|c|}
\hline
$p$ & $\log \mc{Z}^p_{\rm bulk, fin}$ & $\log \mc{Z}^p_{\rm edge, fin}$ & $\log \mc{Z}^p_{\rm PI, fin}=\log \mc{Z}^p_{\rm bulk, fin}+\log \mc{Z}^p_{\rm edge, fin}$  \\
\hline
0 & $- \frac{23 \zeta (3)}{48 \pi^2}+\frac{\zeta (5)}{16 \pi^4} $ & $\log \frac{2\pi^{3/2} R^{3/2}}{q}$ & $\log \frac{2\pi^{3/2} R^{3/2}}{q}- \frac{23 \zeta (3)}{48 \pi^2} + \frac{\zeta (5)}{16 \pi^4} $ \\
\hline
1 & $\frac{\zeta (3)}{16 \pi^2} + \frac{3 \zeta (5)}{16 \pi^4}$ & $\log\frac{q}{2\pi \sqrt{R}} + \frac{\zeta (3)}{4 \pi^2}$ & $\log\frac{q}{2\pi \sqrt{R}} + \frac{5 \zeta (3)}{16 \pi^2}+ \frac{3 \zeta (5)}{16 \pi^4}  $ \\
\hline
2 & $\frac{\zeta (3)}{16 \pi^2} + \frac{3 \zeta (5)}{16 \pi^4}$ & $\log \frac{1}{q \sqrt{R}}+ \frac{\zeta (3)}{4 \pi^2}$ & $\log \frac{1}{q \sqrt{R}} + \frac{5 \zeta (3)}{16 \pi^2}+ \frac{3 \zeta (5)}{16 \pi^4} $ \\
\hline
3 & $- \frac{23 \zeta (3)}{48 \pi^2}+\frac{\zeta (5)}{16 \pi^4} $ & $\log\left(q R^{3/2} \pi^{1/2}\right)$ & $\log\left(q R^{3/2} \pi^{1/2}\right)- \frac{23 \zeta (3)}{48 \pi^2} + \frac{\zeta (5)}{16 \pi^4} $ \\
\hline
\end{tabular}
\caption{$D=5$}
\label{tab:gammaD=5}
\end{table}

\begin{table}[H]
\centering
\renewcommand{\arraystretch}{1.5} 
\begin{tabular}{|c|c|c|c|}
\hline
$p$ & $\log \mc{Z}^p_{\rm bulk, fin}$ & $\log \mc{Z}^p_{\rm edge, fin}$ & $\log \mc{Z}^p_{\rm PI, fin}=\log \mc{Z}^p_{\rm bulk, fin}+\log \mc{Z}^p_{\rm edge, fin}$  \\
\hline
0 & $-\frac{949\,\zeta(3)}{1440\,\pi^2} + \frac{13\,\zeta(5)}{48\,\pi^4} - \frac{\zeta(7)}{64\,\pi^6}$ & $\log\frac{\sqrt{2}\,\pi^2\,R^{5/2}}{q}$ & $\log\frac{\sqrt{2}\,\pi^2\,R^{5/2}}{q} - \frac{949\,\zeta(3)}{1440\,\pi^2}+ \frac{13\,\zeta(5)}{48\,\pi^4}  - \frac{\zeta(7)}{64\,\pi^6}$ \\
\hline
1 & $\frac{41\,\zeta(3)}{288\,\pi^2} + \frac{5\,\zeta(5)}{12\,\pi^4} - \frac{5\,\zeta(7)}{64\,\pi^6}$ & $\log\frac{q}{2\,\pi^{3/2}\,R^{3/2}} + \frac{23\,\zeta(3)}{48\,\pi^2}- \frac{\zeta(5)}{16\,\pi^4} $ & $\log\frac{q}{2\,\pi^{3/2}\,R^{3/2}} + \frac{179\,\zeta(3)}{288\,\pi^2} + \frac{17\,\zeta(5)}{48\,\pi^4}  - \frac{5\,\zeta(7)}{64\,\pi^6}$ \\
\hline
2 & $-\frac{\zeta(3)}{36\,\pi^2} - \frac{5\,\zeta(5)}{48\,\pi^4} - \frac{5\,\zeta(7)}{32\,\pi^6}$ & $\log\frac{2\,\pi\,R^{1/2}}{q}- \frac{5\,\zeta(3)}{16\,\pi^2}- \frac{3\,\zeta(5)}{16\,\pi^4} $ & $\log\frac{2\,\pi\,R^{1/2}}{q} - \frac{49\,\zeta(3)}{144\,\pi^2} - \frac{7\,\zeta(5)}{24\,\pi^4} - \frac{5\,\zeta(7)}{32\,\pi^6}$ \\
\hline
3 & $-\frac{\zeta(3)}{36\,\pi^2} - \frac{5\,\zeta(5)}{48\,\pi^4} - \frac{5\,\zeta(7)}{32\,\pi^6}$ & $\log q \sqrt{R}  - \frac{5\,\zeta(3)}{16\,\pi^2}- \frac{3\,\zeta(5)}{16\,\pi^4}$ & $\log q \sqrt{R}- \frac{49\,\zeta(3)}{144\,\pi^2}  - \frac{7\,\zeta(5)}{24\,\pi^4} - \frac{5\,\zeta(7)}{32\,\pi^6}$ \\
\hline
4 & $\frac{41\,\zeta(3)}{288\,\pi^2} + \frac{5\,\zeta(5)}{12\,\pi^4} - \frac{5\,\zeta(7)}{64\,\pi^6}$ & $\log\frac{1}{\pi^{1/2}\,R^{3/2}\,q} + \frac{23\,\zeta(3)}{48\,\pi^2}- \frac{\zeta(5)}{16\,\pi^4} $ & $\log\frac{1}{\pi^{1/2}\,R^{3/2}\,q} + \frac{179\,\zeta(3)}{288\,\pi^2}+ \frac{17\,\zeta(5)}{48\,\pi^4}  - \frac{5\,\zeta(7)}{64\,\pi^6}$ \\
\hline
5 & $-\frac{949\,\zeta(3)}{1440\,\pi^2} + \frac{13\,\zeta(5)}{48\,\pi^4} - \frac{\zeta(7)}{64\,\pi^6}$ & $\log\frac{q\,\pi\,R^{5/2}}{\sqrt{2}}$ & $\log\frac{q\,\pi\,R^{5/2}}{\sqrt{2}} - \frac{949\,\zeta(3)}{1440\,\pi^2}+ \frac{13\,\zeta(5)}{48\,\pi^4}  - \frac{\zeta(7)}{64\,\pi^6}$ \\
\hline
\end{tabular}
\caption{$D=7$}
\label{tab:gammaD=7}
\end{table}
\noindent Another case of general interest is a $U(1)$ 3-form gauge field in $D=11$:
\begin{align}
    \log \mc{Z}^{p=3}_{\rm bulk, fin} [dS_{11}] & =   \frac{193 \zeta (3)}{4800 \pi ^2}+\frac{1877 \zeta
   (5)}{11520 \pi ^4}+\frac{217 \zeta (7)}{640 \pi
   ^6}+\frac{77 \zeta (9)}{256 \pi ^8}-\frac{21
   \zeta (11)}{256 \pi ^{10}}\nn\\
    \log \mc{Z}^{p=3}_{\rm edge, fin} [S^{11}] & = -\log \mc{Z}^{p=2}_{\rm PI, fin} [S^{9}]  = \log \frac{q}{2 \sqrt{\pi^3 R^3}} +\frac{1987 \zeta (3)}{2880 \pi ^2} +\frac{2333 \zeta (5)}{3840 \pi
   ^4}+\frac{71 \zeta (7)}{256
   \pi ^6}-\frac{21
   \zeta (9)}{256 \pi ^8}\nn\\
    \log \mc{Z}^{p=3}_{\rm PI, fin} [S^{11}] & =  \log \frac{q}{2\sqrt{\pi^3 R^3}}+\frac{5257 \zeta (3)}{7200 \pi
   ^2}+\frac{2219 \zeta (5)}{2880 \pi
   ^4}+\frac{789 \zeta (7)}{1280 \pi
   ^6}+\frac{7 \zeta
   (9)}{32 \pi ^8}-\frac{21
   \zeta (11)}{256 \pi ^{10}} \;. 
\end{align}
We remark that the last columns of tables \ref{tab:gammaD=3}-\ref{tab:gammaD=7} have appeared in table 2 of \cite{David:2021wrw}, which, however, have not taken into account the first $q$-dependent terms.\footnote{We have also corrected some sign errors in table 2 of \cite{David:2021wrw}.} We also note that because of the absence of $\mu$-dependence in odd $D$, the fundamental charge $q$ and the radius $R$ of the sphere must enter through the dimensionless combination $q R^{p-\frac{D-2}{2}}$.

Because of the duality invariance of the characters \eqref{eq:HCduality}, we trivially have $\log \mc{Z}^p_{\rm bulk}|_{\rm fin}=\log \mc{Z}^{D-2-p}_{\rm bulk}|_{\rm fin}$. With the dual identification $\tilde q = 2\pi/q$ we have precisely $\log \mc{Z}^p_{\rm edge, fin}=\log \mc{Z}^{D-2-p}_{\rm edge,fin}$ and $\log \mc{Z}^p_{\rm PI,fin}=\log \mc{Z}^{D-2-p}_{\rm PI,fin}$, in exact agreement with the general results of \cite{Bastianelli:2005vk,Donnelly:2016mlc}.


\subsection{Entanglement entropy for even spheres and the conformal case}
\label{sec:ConfCase}

A standard path integral argument suggests that the entanglement entropy of the Hartle-Hawking state on global $dS_D$, across the bifurcation surface $S^{D-2}$ of the Cauchy slice $S^{D-1}$, is given by the thermal entropy of the de Sitter static patch. In \cite{Dowker:2017flz}, de Sitter Renyi entropies for massless $p$-forms were computed in terms of partition functions $\mc{Z}^p_{\rm PI} \left[S_\beta^D\right]$ on branched spheres $S^D_\beta$ using zeta function techniques, with $\beta$ controlling the conical deficit (the round sphere case corresponds to $\beta=2\pi$).\footnote{The parameter $q$ in \cite{Dowker:2017flz} should be identified with $\frac{\beta}{2\pi}$ here.} The entanglement entropy was then obtained by setting $\beta=2\pi$. With explicit computations for a few $p$ and even $D$, a pattern was found in \cite{Dowker:2017flz} that the coefficient of the logarithmic divergence of the entanglement entropy was simply that of $\log \mc{Z}^p_{\rm PI}\left[S^D\right] $, i.e.
\be S_{\rm EE, log}= \left(1-\beta\p_\beta \right) \log \mc{Z}_{\rm PI,log}\left[S_\beta^D\right] \big|_{\beta=2\pi} = \log \mc{Z}_{\rm PI,log}\left[S^D\right] \; . \label{eq:EElogZlog}\ee
We will (partially) prove this for any $p$ and even $D$ below. Granting this, the bulk-edge split \eqref{eq:ZDEMsplitSD} of the brick-wall regulated DEM partition function implies a bulk-edge split for the entanglement entropy as well. This is in accordance with earlier calculations of entanglement entropy \cite{Donnelly:2016mlc, Dowker:2017flz, Moitra:2018lxn}, which suggested the existence of edge contributions from a $(p-1)$-form residing on the codimension-2 entangling surface. Our findings corroborate these earlier observations and offer a dynamical framework to understand them as large gauge transformations and their symplectic conjugates, captured by the DEM boundary condition \eqref{eq:DEMBC}.

\paragraph{(Attempted) proof of \eqref{eq:EElogZlog}}

To begin, we note that the path integral for massive or massless $p$-forms on $S^D_\beta$ can be put into the form 
\begin{align}\label{eq:masslessPIbulkedgebranch}
    \mc{Z}^{p}_{\rm PI} \left[S_\beta^D\right] = \mc{Z}^p_{\rm bulk}(\beta)\mc{Z}^p_{\rm edge}\left[S_\beta^D\right] \; .
\end{align}
The bulk part is given by the quasicanonical partition function at inverse temperature $\beta $ 
(we suppress the factor $\mu$ for notational simplicity)
\begin{align}\label{eq:Zbulkbranch}
    \log \mc{Z}^p_{\rm bulk} (\beta)= \int_0^\infty \frac{dt}{2t}\frac{1+e^{-\frac{2\pi }{\beta  R}t}}{1-e^{-\frac{2\pi}{\beta R}t}} \chi \left( \frac{t}{ R}\right) \; ,
\end{align}
where $\chi( t)$ the Harish-Chandra character of the massive or massless $p$-form. The edge partition function $\mc{Z}^p_{\rm edge}\left[S_\beta^D\right]$ is
independent of $\beta$, so without loss of generality we can set $\beta=2\pi$ for it, reducing to \eqref{eq:ZedgeSd-2} in the massless case and \eqref{appeq:Zedgemassivepform} in the massive case. The formula \eqref{eq:masslessPIbulkedgebranch} was proposed in \cite{David:2021wrw} and can be proved more generally \cite{Law:2025ktz}.

The fact that only the bulk part
\eqref{eq:Zbulkbranch}
is sensitive to $\beta$ implies that \eqref{eq:EElogZlog} is equivalent to 
\begin{align}\label{eq:bulkder}
    \partial_\beta \log \mc{Z}^p_{\rm bulk,log}\left(\beta\right) |_{\beta =2\pi} =0 \;.
\end{align}
To proceed, consider first the case of a $p$-form with generic mass $m^2$, with Harish-Chandra character
\begin{align}\label{eq:scalarHC}
    \chi(t) = \binom{D-1}{p}\frac{e^{-\Delta t}+e^{-\bar\Delta t}}{\left( 1-e^{-t}\right)^{D-1}} \; , \qquad m^2 R^2 =\left(\Delta-p\right) \left(\bar\Delta-p\right)=\left(\Delta-p\right)\left(D-1-\Delta-p\right) \;. 
\end{align}
With the formula \eqref{eq:Zbulkbranch}, we can deduce 
\begin{align}\label{eq:ZbulkDscalar}
    \beta \partial_\beta\log \mc{Z}^{p}_{\rm bulk}\left(\beta\right)|_{\beta=2\pi} = \int_0^\infty \frac{dt}{R} \binom{D-1}{p} \frac{e^{-(\Delta+1) \frac{t}{R}}+e^{-(\bar\Delta+1) \frac{t}{R}}}{\left( 1-e^{-\frac{t}{R}}\right)^{D+1}} \; . 
\end{align}
The log divergence of \eqref{eq:ZbulkDscalar} is given by the coefficient of $\frac{1}{t}$ in the small-$t$ expansion of the integrand (see section \ref{sec:SDnumerical}). We find the following remarkably simple formula,
\begin{align}\label{eq:logformula}
    \frac{1}{R} \frac{e^{-(\Delta+1) \frac{t}{R}}+e^{-(\bar\Delta+1) \frac{t}{R}}}{\left( 1-e^{-\frac{t}{R}}\right)^{D+1}} = \cdots + 2 \binom{\Delta}{D}  \frac{1}{t}+\cdots \;.
\end{align}
We checked this up to $D=100$, but we do not have an analytic derivation. With \eqref{eq:logformula}, we have
\begin{align}\label{eq:Zbulkderlog}
    \beta \partial_\beta\log \mc{Z}^{p}_{\rm bulk,log}\left(\beta\right)|_{\beta=2\pi} = \frac{2}{D!} \frac{\Gamma(\Delta+1)}{\Gamma(\Delta+1-D)} \binom{D-1}{p}\log \mu R \;. 
\end{align}
This is non-zero for generic values of $\Delta$, implying that \eqref{eq:EElogZlog} does not hold for $p$-forms with generic masses. On the other hand, the Harish-Chandra characters \eqref{eq:HCcharpgauge} for massless $p$-forms involve terms of the form \eqref{eq:scalarHC} with
\begin{align}
     \Delta =0 , 1, \cdots , p\leq D-1 \;.
\end{align}
For example, the $k=0$ term of \eqref{eq:HCcharpgauge} takes the form \eqref{eq:scalarHC} with $\Delta=0$. For these values of $\Delta$, \eqref{eq:Zbulkderlog} vanishes. Note that the $-(-)^p$ term in \eqref{eq:HCcharpgauge} does not alter the coefficient of $\frac{1}{t}$ in the small-$t$ expansion. This establishes the relation \eqref{eq:EElogZlog} (up to a proof of \eqref{eq:logformula}).

\paragraph{Conformal $p$-forms}

When $D$ is even, the $p$-form gauge theory is conformal when $p = \frac{D-2}{2}$, and in this case the de Sitter entanglement entropy is also the entanglement across a sphere of radius $R$ in flat space. In even-dimensional CFTs, the universal coefficient of $\log \mu R$ in the entanglement entropy coincides with that of the sphere partition function, which is in turn determined by the type A conformal anomaly \cite{Casini:2011kv}. In principle the entanglement entropy in CFT can equally well be computed in terms of the partition function on a hyperbolic cylinder $S_\beta^1\times AdS_{D-1}$, but several authors' attempts to do this for $p=1$ in $D=4$, $p=2$ in $D=6$ gave answers that disagreed with the conformal anomaly \cite{Huang:2014pfa, Nian:2015xky, Beccaria:2017dmw, David:2020mls}. It was later shown for any $p$ in \cite{David:2021wrw} that the hyperbolic cylinder method (with standard boundary conditions) always reproduces only the bulk part of the sphere partition function. Our results provide a physical explanation for the missing edge piece.

\subsection{A Donnelly-Wall-style prescription for $p$-forms}
\label{sec:DW}

In the context of Maxwell theory, the edge partition function was first obtained in \cite{Donnelly:2015hxa}, drawing inspiration from lattice gauge theory. The edge path integral was introduced as an integral over superselection sectors labeled by electric flux through the boundary, and was designed with shrinkability in mind. While this proposal resolved some puzzles \cite{Dowker:2010bu, Casini:2011kv, Eling:2013aqa, Huang:2014pfa}, it relied on ad hoc assumptions. Indeed, understanding the underlying dynamical principles was part of the motivation of \cite{Ball:2024hqe}  and the current work. In \cite{Ball:2024hqe}, we provided a detailed comparison between our dynamical framework and the prescription of \cite{Donnelly:2015hxa}, and offered a straightforward generalization to the massive vector case.

In this subsection we extend the prescription of \cite{Donnelly:2015hxa} to general massless $p$-form gauge theories, noting that the correct measure for $E_\perp$ deviates from the na\"ive lattice-based expectation. We also extend it to massive $p$-forms, which involve additional subtleties compared to the $p=1$ case.

\subsubsection{Massive $p$-forms}

We first study a massive $p$-form on the Euclidean manifold $\mc{M}$ considered in section \ref{sec:shrink}, namely one obtained from the static Lorentzian manifold $M$ with metric \eqref{eq:staticmetrichorizon} by a Wick rotation $t\to-i\tau$ with $\tau \sim \tau +2\pi$. Following section \ref{sec:shrink}, we once again cut out a hole of proper radius $\eps$ encircling the origin, resulting in a manifold $\mc{M}_\eps$. Following \cite{Donnelly:2015hxa}, we would like to compute the partition function\footnote{Classically, a massive $p$-form is dual to a massive $(D-1-p)$-form. See for instance \cite{Buchbinder:2008jf,Kuzenko:2020zad} for discussions on their quantum equivalence.}
\begin{gather}\label{eq:pformPI}
    Z^{p,m^2}_{\rm PI} [E_\perp, \mc{M}_\eps]= \int \mathcal{D}A \, e^{-S[A]} \; , \quad 
    S = \half \int_{\mc{M}_\eps} \left( F \wedge *_{\mc{M}_\eps} F + m^2 A \wedge *_{\mc{M}_\eps} A \right)
\end{gather}
with the PMC-like boundary condition:\footnote{Technically one should add a boundary term to the action to make this boundary condition variationally well-defined. This was ignored in \cite{Donnelly:2015hxa}, but it just changes the action by an overall sign that is canceled anyway by another sign from using the Euclidean rather than Lorentzian electric field. We thank Hong Zhe Chen for discussions on this point.}
\begin{align}\label{eq:PMClike}
    n^\nu F_{\tau \nu a_1\dots a_{p-1}}|_{\p \mc{M}_\eps} = -\sqrt{g_{\tau\tau}}E_{\perp, a_1\dots a_{p-1}} \; , \qquad n^\nu F_{\nu i_1\dots i_p}|_{\p \mc{M}_\eps} =0 \; . 
\end{align}
Here $n^\nu$ is the outward unit normal at $\p \mc{M}_\eps$. The $x^a$ are coordinates on $\p\Sigma$. The $E_{\perp}$ in the first equation is a prescribed boundary normal component of the ``electric" field, defined as
\be E_{i_1\dots i_p} \equiv -\sqrt{g^{\tau\tau}} F_{\tau i_1\dots i_p} \; , \qquad  E_{\perp,a_1\dots a_{p-1}} \equiv n^i E_{i a_1\dots a_{p-1}} \; . \ee
To proceed, we separate the $p$-forms into a classical and a fluctuation part
\begin{align}\label{eq:separate}
    A_{\mu_1 \cdots \mu_{p}} = A_{\mu_1 \cdots \mu_{p}}^{\rm cl}[E_\perp] + A^{\rm fluc}_{\mu_1 \cdots \mu_{p}}
\end{align}
where $A_{\mu_1 \cdots \mu_{p}}^{\rm cl}[E_\perp]$ is the unique solution to the equation of motion
\begin{align}\label{eq:procaeom}
     \nabla^\lambda F_{\lambda \mu_1 \cdots \mu_{p}}= m^2 A_{\mu_1 \cdots \mu_{p}}
\end{align}
subject to \eqref{eq:PMClike}, while the fluctuation $A^{\rm fluc}_{\mu_1 \cdots \mu_{p}}$ obeys the PMC boundary condition
\begin{align}\label{eq:PMCexact}
    n^\nu F^{\rm fluc}_{\nu\mu_1\dots\mu_p}|_{\p \mc{M}_\eps} =0 \; . 
\end{align}
Since the action \eqref{eq:pformPI} is quadratic, upon the separation \eqref{eq:separate} the path integral splits into
\begin{align}
    Z^{p,m^2}_{\rm PI} [ E_\perp, \mc{M}_\eps ] = 
    e^{-S_{\text{on-shell}}[E_\perp]}Z^{p,m^2}_{\rm  bulk} [\mc{M}_\eps ] \; . 
\end{align}
Here the path integral $Z^{p,m^2}_{\rm  bulk} [\mc{M}_\eps ]$ captures the integrations over the bulk fluctuation $A^{\rm fluc}$ obeying  \eqref{eq:PMCexact}. The on-shell action reduces to a boundary term
\begin{align}\label{eq:Sonshell}
    S_{\text{on-shell}}[E_\perp] = \frac{1}{2p!} \int_{\p \mc{M}_\eps} A_{\mu_1 \cdots \mu_p} n_\nu F^{\nu \mu_1 \cdots \mu_{p}} \; .
\end{align}
Mimicking \cite{Donnelly:2015hxa}, we propose an edge path integral
\begin{align}\label{eq:procaDWedgePI}
    Z^{p,m^2}_{\rm edge} [\mc{M}_\eps ] \equiv \int \mathcal{D} E_\perp \, e^{-S_{\text{on-shell}}[E_\perp]} \; ,
\end{align}
in which we only include $\tau$-independent $E_\perp$ configurations. 

\paragraph{Calculation of $Z^{p,m^2}_{\rm edge}[\mc{M}_\eps]$}

To compute \eqref{eq:procaDWedgePI}, we need to find the unique solution to the equation of motion subject to \eqref{eq:PMClike} for any given $E_\perp$ on $\p \mc{M}_\eps$. We make the ansatz that the only nonzero components are those with a $\tau$ index, and that they are independent of $\tau$. This implies that 
\be E_{i_1\dots i_p} = \sqrt{g^{\tau\tau}} p! \, \hat\nabla_{[i_1} \gam_{i_2\dots i_p]} \ee
where $\hat\nabla_i$ is the covariant derivative with respect to the metric $g_{ij}$ on $\Sigma$, and we have defined
\be \gam_{i_1\dots i_{p-1}} \equiv A_{\tau i_1\dots i_{p-1}} \; . \ee
This $(p-1)$-form turns out to play a highly analogous role to the edge mode $\beta$ in the $p$-form gauge theory in section \ref{sec:CovPhaseEdge}. Imposing the equation of motion \eqref{eq:procaeom}, it obeys
\be \hat\nabla^j (\sqrt{g^{\tau\tau}} \gam_{ji_1\dots i_{p-2}}) = 0 \; , \qquad m^2 \gam_{i_1\dots i_{p-1}} = \frac{1}{\sqrt{g^{\tau\tau}}} \hat\nabla^j \left(\sqrt{g^{\tau\tau}} p! \hat\nabla_{[j} \gam_{i_1\dots i_{p-1}]}\right) \;. \ee
Aside from the $m^2$ term, these are the same equations satisfied by $\beta$ in \eqref{eq:beta-eqns}. The on-shell action \eqref{eq:Sonshell} reduces to
\be S[E_\perp] = \frac{2\pi}{2(p-1)!} \int_{\p\Sigma_\eps} \gam_{a_1\dots a_{p-1}} E_\perp^{a_1\dots a_{p-1}} \ee
where $2\pi$ comes from the coordinate periodicity of $\tau$. Recall we are normalizing $\tau$ such that $\kappa=1$.

It remains to solve for $\gam|_{\p\Sigma_\eps}$ in terms of $E_\perp$ in the limit $\eps \to 0$. This problem is analogous to inverting the Dirichlet-to-Neumann operator $K$ in the horizon limit in section \ref{sec:horlim}. One can show that $\gam$ has the same asymptotic behavior near the boundary that $\beta$ did, except for the replacement $\Delta^{\p\Sigma_\eps} \to \Delta^{\p\Sigma_\eps} + m^2$. That is, for small $\eps$ we have
\be \sqrt{g^{\tau\tau}} n^i \p_i \approx \log(\eps^{-1}) \left( \Delta^{\p\Sigma_\eps} + m^2 \right) \; . \label{eq:normalderapp}\ee
Note also that the equation of motion determines $\gam$'s boundary normal component in terms of $E_\perp$. Explicitly we have
\be m^2 \sqrt{g^{\tau\tau}} \textbf{i}_n \gam = \bar d^\dag E_\perp \ee
where we have introduced the exterior derivative $\bar d$ on $\p\Sigma_\eps$ and its adjoint $\bar d^\dag$, as well as the notation $\bf{i}_n$ for plugging the unit normal vector $n^i$ into the first slot of a differential form. With this notation we can compactly write
\be \ba E_\perp & = \textbf{i}_n E  = \sqrt{g^{\tau\tau}} \p_n \gam - \sqrt{g^{\tau\tau}} \bar d \textbf{i}_n \gam \approx \sqrt{g^{\tau\tau}} \p_n \gam - \frac{1}{m^2} \bar d \, \bar d^\dag E_\perp \; . \ea \ee
In the last step we have plugged in for $\bf{i}_n\gam$ and made use of the fact that $\sqrt{g_{\tau\tau}}$ and $\bar d$ approximately commute in the $\eps\to 0$ limit. Rearranging slightly and plugging in the approximate form \eqref{eq:normalderapp} for $\sqrt{g^{\tau\tau}}\p_n$, we have
\be \left( 1 + \frac{1}{m^2} \bar d \, \bar d^\dag \right) E_\perp \approx \log(\eps^{-1}) \left(\Delta^{\p\Sigma_\eps} + m^2\right) \gam \; . \ee
Next we decompose $E_\perp$ into co-closed and exact parts,
\be E_\perp = E_\perp^T + \bar d\chi \; , \ee
where $E_\perp^T$ is co-closed. We also choose $\chi$ to be co-exact so that the decomposition is unique. The co-closed part of $E_\perp$ satisfies
\be E_\perp^T = \log(\eps^{-1}) \left( \Delta^{\p\Sigma_\eps} + m^2 \right) \gam \; . \ee
Recalling that $\Delta^{\p\Sigma_\eps} = \bar d \, \bar d^\dag + \bar d^\dag \bar d$, we see that the exact part satisfies
\be \frac{1}{m^2} \left(  \Delta^{\p\Sigma_\eps} +m^2 \right) \bar d\chi = \log(\eps^{-1}) \left( \Delta^{\p\Sigma_\eps} + m^2 \right) \gam \; . \ee
This means that $\gam = \frac{\bar d\chi}{m^2 \log(\eps^{-1})}$. When we plug into the action it splits into two pieces, $S[E_\perp] = S[E_\perp^T] + S[\bar d\chi]$. The first is
\be S[E_\perp^T] = \frac{2\pi}{2\log(\eps^{-1})} \int_{\p\Sigma_\eps} E^T_\perp \wedge *_{\p\Sigma_\eps} \frac{1}{\Delta^{\p\Sigma_\eps} + m^2} E_\perp^T \; . \ee
The second piece takes the form
\be S[\bar d\chi] = \frac{2\pi}{2m^2\log(\eps^{-1})} \int_{\p\Sigma_\eps} \bar d\chi \wedge *_{\p\Sigma_\eps} \bar d\chi = \frac{2\pi}{2m^2\log(\eps^{-1})} \int_{\p\Sigma_\eps} \chi \wedge *_{\p\Sigma_\eps} \Delta^{\p\Sigma_\eps} \chi \; , \ee
where we used $\bar d^\dag \chi=0$ to simplify. Our change of variables produces the Jacobian
\be \mc{D} E_\perp = \mc{D} E_\perp^T \mc{D}(\bar d\chi) = \mc{D} E_\perp^T \mc{D}\chi \det\hspace{0mm}'(\Delta^{T,\p\Sigma_\eps}_{p-2})^{1/2} \; . \ee
Finally, path integrating gives
\be \ba Z^{p,m^2}_{\rm edge}[\mc{M}_\eps] & = \det\hspace{0mm}'\left(\Delta^{T,\p\Sigma_\eps}_{p-2}\right)^{1/2} \int \mc{D}E_\perp^T \int \mc{D}\chi \, e^{-S[E_\perp^T] - S[\bar d\chi]} \\
& = \det\hspace{0mm}'\left(\Delta^{T,\p\Sigma_\eps}_{p-2}\right)^{1/2} \det(\frac{\log(\eps^{-1})}{2\pi\mu^2} (\Delta^{T,\p\Sigma_\eps}_{p-1} + m^2))^{1/2} \det\hspace{0mm}'\left( \frac{m^2\log(\eps^{-1})}{2\pi\mu^2} \Delta^{T,\p\Sigma_\eps}_{p-2}\right)^{-1/2} \; . \ea \ee
As in the discussion after \eqref{eq:Zbedge-penul}, the factor of $\frac{m^2\log(\eps^{-1})}{2\pi\mu^2}$ in the latter determinant can be pulled out with some power involving Betti numbers and an anomaly. The massless determinants then cancel. The factor of $\frac{\log\eps^{-1}}{2\pi}$ in the massive determinant can also be pulled out with some power. Overall we have
\be \label{eq:DWmassiveresult} Z^{p,m^2}_{\rm edge}[\mc{M}_\eps] = \left( \frac{m^2\log(\eps^{-1})}{2\pi\mu^2} \right)^{\mc{A} - \half \sum_{k=0}^{p-2} (-)^k b_k} \left( \frac{\log(\eps^{-1})}{2\pi} \right)^{\mc{A}'} \det(\mu^{-2}(\Delta^{T,\p\Sigma_\eps}_{p-1} + m^2))^{1/2} \ee
where $\mc{A}$, $\mc{A}'$ are some in principle computable anomalies present only when $D$ is even. Up to the constant factors out front, we have the (reciprocal) partition function of a massive $(p-1)$-form on $\p\Sigma_\eps$. To summarize, the object
\be \ba
   Z_{\rm DW}^{p,m^2}[\mc{M}_\eps] & \equiv \int \mathcal{D} E_\perp \, Z^{p,m^2}_{\rm PI} [ E_\perp, \mc{M}_\eps ] \\
   & = Z^{p,m^2}_{\rm  bulk} [\mc{M}_\eps ]   Z^{p,m^2}_{\rm edge}[\mc{M}_\eps]
\ea \ee
splits into bulk and edge parts and is supposed to be shrinkable, recovering the partition function on the closed manifold $\mc{M}$ in the $\eps\to 0$ limit. To contextualize this result, let us consider the case $\mc{M}=S^D$, where we have demonstrated the shrinkability of our DEM boundary condition for massless $p$-form gauge theories in section \ref{sec:shrink}. Similar to the massless case discussed in section \ref{sec:ZPISDnohole}, the partition function for a massive $p$-form on a round $S^D$ with radius $R$ exhibits a bulk-edge split (reviewed in appendix \ref{app:massivepformSD}) \cite{Anninos:2020hfj}
\begin{align}\label{eq:massveSDsplit}
	\mc{Z}^{p,m^2}_{\rm PI} \left[S^D \right] =& \mc{Z}^{p,m^2}_{\rm bulk}\left[dS_D\right] \mc{Z}^{p,m^2}_{\rm edge} \left[S^{D} \right] \;. 
\end{align}
The bulk part is given by a quasicanonical partition function at inverse temperature $\beta = 2\pi$
\begin{align}\label{eq:massiveZbulk}
	\log \mc{Z}^{p,m^2}_{\rm bulk}\left[dS_D\right]  = & \int_0^\infty \frac{dt}{2t} \frac{1+e^{-\frac{t}{R}}}{1-e^{-\frac{t}{R}}}\chi \left( \frac{t}{R}\right) \; , \qquad \chi(t) = \binom{D-1}{p}\frac{e^{-\Delta t} +e^{-\bar\Delta t}}{\left|1-e^{-t} \right|^{D-1}}  \;, 
\end{align}
defined in terms of the Harish-Chandra character $\chi(t)$ for a $p$-form with mass $m^2R^2=\left(\Delta-p\right) \left(\bar\Delta-p\right) = \left(\Delta-p\right)\left(D-1-\Delta-p\right)$. The edge partition function $\mc{Z}^{p,m^2}_{\rm edge}$ is the reciprocal of a $(p-1)$-form partition function with mass $m^2$ on $S^{D-2}$: 
\begin{align}\label{eq:Zedgeproca}
	 \mc{Z}^{p,m^2}_{\rm edge} \left[S^{D} \right] =  \frac{1}{\mc{Z}^{p-1,m^2}_{\rm PI} \left[S^{D-2} \right]}\; . 
\end{align}
In \cite{Ball:2024hqe}, we found for the $p=1$ case that neither the symplectic form nor the Hamiltonian split into a bulk and edge piece as in the massless case. Nonetheless, the edge partition function defined with the prescription \eqref{eq:procaDWedgePI} reproduced the known result of the edge partition function in the case when $\mc{M}=S^D$, namely that the edge partition function is the reciprocal of a scalar partition function with mass $m^2$ on $S^{D-2}$ \cite{Anninos:2020hfj}. Comparing \eqref{eq:DWmassiveresult} with \eqref{eq:Zedgeproca}, we can see that this result extends to general $p\geq 1$: the edge mode partition function defined with the prescription \eqref{eq:procaDWedgePI} reproduces the edge partition function \eqref{eq:Zedgeproca} (up to the constant factors in \eqref{eq:DWmassiveresult}).

\subsubsection{$p$-form gauge theory}

We now carry out a similar analysis for $p$-form gauge theory. For simplicity we assume $\Sigma_\eps$ has the topology of a $(D-1)$-ball and that $2 \le p \le D-2$. Once again we modify the PMC boundary condition by specifying a $\tau$-independent electric flux on the boundary. Now the equation of motion requires it to be co-closed, so we write it as $E^T_\perp$. The partition function defined by this boundary condition factorizes into classical and fluctuation parts,
\be Z_{\rm PI}^p[E_\perp^T, \mc{M}_\eps] = e^{-S_{\text{on-shell}}[E^T_\perp]} Z^p_{\rm  bulk} [\mc{M}_\eps] \; , \ee
where $Z_{\rm bulk}^p[\mc{M}_\eps]$ uses the PMC boundary condition. We prescribe the full partition function
\be \ba Z_{\rm DW}^p[\mc{M}_\eps] & \equiv \frac{|\mc{C}^{p-1}(\p\Sigma_\eps)|}{|\mc{G}^{{\p\Sigma_\eps}}|} \int \mc{D}E^T_\perp \, Z^{p}_{\rm PI}[E^T_\perp, \mc{M}_\eps] \\
& = \frac{|\mc{C}^{p-1}(\p\Sigma_\eps)|}{|\mc{G}^{{\p\Sigma_\eps}}|} Z^{p}_{\rm bulk}[\mc{M}_\eps] \int \mc{D}E^T_\perp \, e^{-S_{\text{on-shell}}[E^T_\perp]} \\
& = Z^{p}_{\rm bulk}[\mc{M}_\eps] Z^{p}_{\rm edge}[\mc{M}_\eps] \label{eq:ZDWPI} \ea \ee
where in the last line we have defined
\be Z_{\rm edge}^p[\mc{M}_\eps] \equiv \frac{|\mc{C}^{p-1}(\p\Sigma_\eps)|}{|\mc{G}^{{\p\Sigma_\eps}}|} \int \mc{D}E^T_\perp \, e^{-S_{\text{on-shell}}[E^T_\perp]} \; . \ee
Here $|\mc{C}^{p-1}(\p\Sigma_\eps)|$ is formally the (infinite) volume of the space of co-closed $(p-1)$-forms on $\mc{M}_\eps$, and the formal quotient reduces to
\be \label{eq:DWmeasure} \frac{|\mc{C}^{p-1}(\p\Sigma_\eps)|}{|\mc{G}^{{\p\Sigma_\eps}}|} = |U(1)|^{(-)^p} \prod_{k=0}^{p-2}\det\hspace{0mm}'\left(\frac{\Delta_k^{T,\p\Sigma}}{\mu^2}\right)^{-\frac{(-)^{p-k}}{2}} \;. \ee
The measure $\frac{|\mc{C}^{p-1}(\p\Sigma_\eps)|}{|\mc{G}^{{\p\Sigma_\eps}}|} \int \mc{D}E^T_\perp$ is an ad hoc choice made so that $Z_{\rm DW}^p[\mc{M}_\eps]$ coincides with our $\bar Z_{\rm DEM}[\beta=2\pi]$. To evaluate the on-shell action $S_{\text{on-shell}}[E^T_\perp]$, we again define
\be \gam_{i_1\dots i_{p-1}} \equiv A_{\tau i_1\dots i_{p-1}} \; , \ee
and the equation of motion implies
\be \hat\nabla^j (\sqrt{g^{\tau\tau}} \gam_{ji_1\dots i_{p-2}}) = 0 \; , \qquad \frac{1}{\sqrt{g^{\tau\tau}}} \hat\nabla^j (\sqrt{g^{\tau\tau}} p! \hat\nabla_{[j} \gam_{i_1\dots i_{p-1}]}) = 0 \;. \ee
It still asymptotically satisfies
\be \sqrt{g^{\tau\tau}} \p_n \gam \approx \log(\eps^{-1}) \Delta^{\p\Sigma_\eps} \gam \; . \ee
The on-shell action reduces to
\be S_{\text{on-shell}}[E^T_\perp] = \frac{2\pi}{2} \int_{\p\Sigma_\eps} \gam \wedge *_{\p\Sigma_\eps} E^T_\perp = \frac{2\pi}{2\log\eps^{-1}} \int_{\p\Sigma_\eps} E^T_\perp \wedge *_{\p\Sigma_\eps} \frac{1}{\Delta^{\p\Sigma_\eps}} E^T_\perp \; . \ee
We then have
\be Z^{p}_{\rm edge}[\mc{M}_\eps] = \frac{|\mc{C}^{p-1}(\p\Sigma_\eps)|}{|\mc{G}^{{\p\Sigma_\eps}}|} \det(\frac{\log\eps^{-1}}{2\pi\mu^2} \Delta^{T,\p\Sigma_\eps}_{p-1})^{1/2} \; . \ee
This precisely matches our $\bar Z_{\rm edge}$ for $\beta=2\pi$. Compare with the first line of \eqref{eq:Zbedge-penul}. The full partition functions then match as well,
\be Z_{\rm DW}^p[\mc{M}_\eps] = \bar Z_{\rm DEM}^p[\mc{M}_\eps] \; , \ee
and so our discussion above in section \ref{sec:shrink} about the $\eps\to 0$ shrinking limit applies equally well to $Z_{\rm DW}^p[\mc{M}_\eps]$. The integral over $E_\perp$ here can be understood in terms of non-dynamical superselection sectors, but the additional ad hoc measure factor \eqref{eq:DWmeasure} seems more difficult to interpret from this perspective.

\section*{Acknowledgments} 

It is a great pleasure to thank Changha Choi, Jaume Gomis, and Jacob McNamara for useful discussions. AB is supported by the Celestial Holography Initiative at the Perimeter Institute for Theoretical Physics and the Simons Collaboration on Celestial Holography. Research at the Perimeter Institute is supported by the Government of Canada through the Department of Innovation, Science and Industry Canada and by the Province of Ontario through the Ministry of Colleges and Universities. AL was supported by the Stanford Science Fellowship and a Simons Investigator award.


\appendix


\section{Conventions for differential forms}
\label{app:DiffConv}

In this appendix we establish our conventions for differential forms on a $D$-dimensional manifold $M$ with metric $g_{\mu\nu}$. We use the antisymmetrization convention that divides by the factorial of the number of indices involved, so for example $\omega_{[\mu\nu]} = \half (\omega_{\mu\nu} - \omega_{\nu\mu})$. We write a $k$-form as
\be \omega = \omega_{\mu_1\dots\mu_k} dx^{\mu_1} \otimes \dots \otimes dx^{\mu_k} = \omega_{\mu_1\dots\mu_k} \frac{1}{k!} dx^{\mu_1} \wedge \dots \wedge dx^{\mu_k} \ee
This determines our wedge product convention. In particular $dx \wedge dy = dx \otimes dy - dy \otimes dx$ and
\be (\omega \wedge \psi)_{\mu_1\dots \mu_{k+\ell}} = \frac{(k+\ell)!}{k! \, \ell!} \omega_{[\mu_1\dots \mu_k} \psi_{\mu_{k+1}\dots\mu_{k+\ell}]}. \ee
The product rule for the exterior derivative then determines
\be (d\omega)_{\mu_1\dots \mu_{k+1}} = (k+1) \p_{[\mu_1} \omega_{\mu_2\dots\mu_{k+1}]}. \ee
We define the volume form $\eps$ such that $\eps_{12\dots D} = \sqrt{|g|}$. The choice of ordering for the coordinates determines the orientation. We define the Hodge dual on $M$ as
\be (*_M\omega)_{\mu_1\dots\mu_{D-k}} \equiv \frac{1}{k!} \, \omega_{\nu_1\dots\nu_k} \eps^{\nu_1\dots\nu_k}{}_{\mu_1\dots\mu_{D-k}}. \ee
It is involutive up to sign. On $k$-forms we have
\be *_M *_M = (-)^{s+k(D-k)} \ee
where $s$ is the signature of the metric, i.e. the number of minus signs in it. We define the inner product
\be (\omega, \psi) \equiv \int_M \omega \wedge *_M\psi = \frac{1}{k!} \int_M d^Dx \sqrt{g} \, \omega_{\mu_1\dots\mu_k} \psi^{\mu_1\dots\mu_k}. \ee
From this we can deduce the adjoint of the exterior derivative. Acting on a $k$-form we have
\be d^\dag = (-)^{s+1+D(k-1)} {*_M d*_M} \; .\label{appeq:dagdef} \ee
In terms of components this is
\be (d^\dag\omega)_{\mu_1\dots\mu_{k-1}} = -\nabla^\nu \omega_{\nu\mu_1\dots\mu_{k-1}}\; . \ee
We often use the identity
\be \nabla_\nu \omega^{\nu\mu_1\dots\mu_{k-1}} = \frac{1}{\sqrt{|g|}} \p_\nu (\sqrt{|g|} \, \omega^{\nu\mu_1\dots\mu_{k-1}}) \; . \ee
The Laplacian is defined as
\be \Delta \equiv (d + d^\dag)^2 = d^\dag d + dd^\dag \; . \label{appeq:Lapdef}\ee
As the square of a self-adjoint operator its spectrum is non-negative. In terms of components its leading term is
\be (\Delta\omega)_{\mu_1\dots\mu_k} = -g^{\nu\rho} \p_\nu \p_\rho \omega_{\mu_1\dots\mu_k} + \text{linear in $\p_\mu$}\; . \ee
We write the pullback of $\omega$ to a submanifold $\Sigma \subset M$ as $i_\Sigma \omega$. We often use the fact that the pullback commutes with the exterior derivative, i.e. $d\, i_\Sigma \omega = i_\Sigma d\omega$.

\section{Horizon limit calculation}
\label{app:horiz-lim}

This appendix carries out the calculation set up in section \ref{sec:horlim}. We solve the equations $d^\ddagger \beta = 0$ and $d^\ddagger d\beta = 0$ asymptotically near the boundary of $\Sigma$ at small $r$. Recall $S = \sqrt{-g^{tt}}$ and $d^\ddagger = \frac{1}{S} d^\dag S$. In (upper) components the equations are respectively
\be \label{appeq:S-c-c-comp} 0 = -\frac{1}{S \sqrt{\hat g}} \p_j \left( S\sqrt{\hat g} \, \beta^{ji_1\dots i_{p-2}} \right) \ee
and
\be 0 =- \frac{p}{S \sqrt{\hat g}} \p_k \left( g^{k\ell} g^{i_1j_1} \dots g^{i_{p-1}j_{p-1}}S\sqrt{\hat g} \, \p_{[\ell} \beta_{j_1\dots j_{p-1}]} \right) \ee
where $\sqrt{\hat g} \equiv \sqrt{\det g_{ij}} = \sqrt{\det g_{ab}}$. The awkward inclusion of many upper metrics in the second expression is the cost of avoiding Christoffel symbols. Our strategy in solving these equations involves splitting the indices along $x^i = (r, x^a)$. When none of the free indices in the latter equation is $r$ we have (with lower indices)
\be \ba 0 & = g_{a_1b_1} \dots g_{a_{p-1}b_{p-1}} \frac{-p}{S\sqrt{\hat g}} \p_i \left( S\sqrt{\hat g} \, g^{ij} g^{b_1c_1} \dots g^{b_{p-1}c_{p-1}} \p_{[j} \beta_{c_1\dots c_{p-1}]} \right) \\
& = g_{a_1b_1} \dots g_{a_{p-1}b_{p-1}} \frac{-p}{S\sqrt{\hat g}} \p_r \left( S\sqrt{\hat g} \, g^{b_1c_1} \dots g^{b_{p-1}c_{p-1}} \p_{[r} \beta_{c_1\dots c_{p-1}]} \right) \\
& \hspace{36mm} + g_{a_1b_1} \dots g_{a_{p-1}b_{p-1}} \frac{-p}{S\sqrt{\hat g}} \p_e \left( S\sqrt{\hat g} \, g^{ef} g^{b_1c_1} \dots g^{b_{p-1}c_{p-1}} \p_{[f} \beta_{c_1\dots c_{p-1}]} \right) \\
& \approx \frac{-p}{S} \p_r \left( S \p_{[r} \beta_{a_1\dots a_{p-1}]} \right) + g_{a_1b_1} \dots g_{a_{p-1}b_{p-1}} \frac{-p}{\sqrt{\hat g}} \p_e \left( \sqrt{\hat g} \, g^{ef} g^{b_1c_1} \dots g^{b_{p-1}c_{p-1}} \p_{[f} \beta_{c_1\dots c_{p-1}]} \right) \; . \ea \ee
In the third line the use of $\,\approx\,$ indicates that we are dropping subleading terms like $\p_r g_{ab}$ and $S^{-1} \p_a S$. The latter term is closely related to the Laplacian on $\p\Sigma$, acting on $(p-1)$-forms,
\be \ba \left((\Delta_{p-1}^{\p\Sigma} \beta\right)_{a_1\dots a_{p-1}} & = g_{a_1b_1} \dots g_{a_{p-1}b_{p-1}} \frac{-p}{\sqrt{\hat g}} \p_e \left( \sqrt{\hat g} \, g^{ef} g^{b_1c_1} \dots g^{b_{p-1}c_{p-1}} \p_{[f} \beta_{c_1\dots c_{p-1}]} \right) \\
& \hspace{25mm} - (p-1) \p_{[a_1} \left( g_{a_2|b_2|} \dots g_{a_{p-1}]b_{p-1}} \frac{1}{\sqrt{\hat g}} \p_c \left( \sqrt{\hat g} \, \beta^{cb_2\dots b_{p-1}} \right) \right) \; . \ea \ee
The vertical bars around $b_2$ indicate that it is omitted from the antisymmetrization, which is only on $a_1, \dots, a_{p-1}$. We can use this to rewrite
\be \ba 0 & \approx \frac{-p}{S} \p_r \left( S \p_{[r} \beta_{a_1\dots a_{p-1}]} \right) + \left(\Delta^{\p\Sigma}_{p-1} \beta\right)_{a_1\dots a_{p-1}} + (p-1) \p_{[a_1} \left( g_{a_2|b_2|} \dots g_{a_{p-1}]b_{p-1}} \frac{1}{\sqrt{\hat g}} \p_c \left( \sqrt{\hat g} \, \beta^{cb_2\dots b_{p-1}} \right) \right) \\
& \approx \frac{-p}{S} \p_r \left( S \p_{[r} \beta_{a_1\dots a_{p-1}]} \right) + \left(\Delta^{\p\Sigma}_{p-1} \beta\right)_{a_1\dots a_{p-1}} + (p-1) \p_{[a_1} \left( g_{a_2|b_2|} \dots g_{a_{p-1}]b_{p-1}} \frac{1}{S\sqrt{\hat g}} \p_c \left( S\sqrt{\hat g} \, \beta^{cb_2\dots b_{p-1}} \right) \right), \ea \ee
In the second line we used $S^{-1} \p_a S \approx 0$ to insert some factors of $S$. The next step is to use \eqref{appeq:S-c-c-comp} to rewrite as
\be \ba 0 & \approx \frac{-p}{S} \p_r \left( S \p_{[r} \beta_{a_1\dots a_{p-1}]} \right) + \left(\Delta^{\p\Sigma}_{p-1} \beta\right)_{a_1\dots a_{p-1}} - (p-1) \p_{[a_1} \left( g_{a_2|b_2|} \dots g_{a_{p-1}]b_{p-1}} \frac{1}{S\sqrt{\hat g}} \p_r \left( S\sqrt{\hat g} \, \beta^{rb_2\dots b_{p-1}} \right) \right) \\
& \approx \frac{-p}{S} \p_r \left( S \p_{[r} \beta_{a_1\dots a_{p-1}]} \right) + \left(\Delta^{\p\Sigma}_{p-1} \beta\right)_{a_1\dots a_{p-1}} - (p-1) \p_{[a_1} \left( \frac{1}{S} \p_r \left( S \beta_{|r|a_2\dots a_{p-1}]} \right) \right) \; . \\
& \approx \frac{-p}{S} \p_r \left( S \p_{[r} \beta_{a_1\dots a_{p-1}]} \right) + \left(\Delta^{\p\Sigma}_{p-1} \beta\right)_{a_1\dots a_{p-1}} - \frac{p-1}{S} \p_r \left( S \p_{[a_1} \beta_{|r|a_2\dots a_{p-1}]} \right) \; . \ea \ee
In the second line we used $\p_r g_{ab} \approx 0$ to bring the many metrics inside the $\p_r$ derivative then used $g_{rr}=1$, and in the third line we used $S^{-1} \p_a S \approx 0$ and the commutativity of partial derivatives to bring the $\p_{[a_1}$ derivative inside. Once again the bars around $r$ indicate that it is not part of the antisymmetrization. Next we expand the first term to get
\be \label{appeq:final-hor} \ba 0 & \approx \frac{-1}{S} \p_r \left( S \p_r \beta_{a_1\dots a_{p-1}} \right) + \frac{p-1}{S} \p_r \left( S \p_{[a_1} \beta_{|r|a_2\dots a_{p-1}]} \right) + \left(\Delta^{\p\Sigma}_{p-1} \beta\right)_{a_1\dots a_{p-1}} - \frac{p-1}{S} \p_r \left( S \p_{[a_1} \beta_{|r|a_2\dots a_{p-1}]} \right) \\
& \approx \frac{-1}{S} \p_r \left( S \p_r \beta_{a_1\dots a_{p-1}} \right) + \left(\Delta^{\p\Sigma}_{p-1} \beta\right)_{a_1\dots a_{p-1}}. \ea \ee
Now that all mixed derivatives have dropped out it is useful to mode expand with respect to $\Delta_{p-1}^{\p\Sigma}$, writing
\be \beta_{a_1\dots a_{p-1}}(r, x^a) = \sum_n \beta_n(r) \omega_{n,a_1\dots a_{p-1}}(x^a). \ee
The modes are orthonormalized,
\be \int_{\p\Sigma} \omega_m \wedge *_{\p\Sigma} \omega_n = \delta_{mn}, \ee
and satisfy
\be \Delta_{p-1}^{\p\Sigma} \omega_n = \lam_n \omega_n \ee
with non-negative eigenvalues $\lam_n$. Projecting our simplified equation \eqref{appeq:final-hor} onto one of these modes gives
\be 0 \approx \frac{-1}{S} \p_r \left(S \p_r \beta_n\right) + \lam_n \beta_n. \ee
Finally plugging in $S = \frac{1}{\kappa r} + \mc{O}(r)$ gives
\be 0 \approx -r \p_r \left(\frac{1}{r} \p_r \beta_n\right) + \lam_n \beta_n. \ee
This is identical to (2.69) from \cite{Ball:2024hqe} for the $p=1$ case.


\section{$dS_D$ static patch and $S^{D}$ partition functions with no brick wall}\label{app:sphere_PI}

\subsection{(Quasi)canonical bulk partition function in $dS_D$ static patch}\label{app:quasicanonical}

An object that plays a prominent role in \cite{Anninos:2020hfj} is the Harish-Chandra character $\chi(g)$, defined as a trace of a group element $g$ over the representation space associated with unitary irreducible representations of the isometry group $SO(1,D)$ of $dS_{D}$. We will be exclusively interested in the character associated with the $dS$ boost, $g=e^{-i \hat H t}$, the generator of the $SO(1,1)$ subgroup.

\paragraph{Massive $p$-form}

The Harish-Chandra character for a massive $p$-form is given by \cite{10.3792/pja/1195522333,10.3792/pja/1195523460,10.3792/pja/1195523378,Basile:2016aen}
\begin{align}\label{eq:charpform}
	\chi^{\Delta,p}(t) = \binom{D-1}{p} \frac{e^{-\Delta t} +e^{-\bar\Delta t}}{\left|1-e^{-t} \right|^{D-1}} \; ,
\end{align}
where the $\mathfrak{so}(1,1)$ weights $\Delta$ and $\bar \Delta\equiv D-1-\Delta$ are related to the mass $m^2$ by 
\begin{align}\label{eq:pformmass}
	m^2 R^2=  \left(\Delta -p  \right)  \left(\bar\Delta -p  \right)  \; .
\end{align}
Here $R$ is the de Sitter length. The overall factor $\binom{D-1}{p}$ is the number of polarizations of the massive $p$-form field.\footnote{For $p = \frac{D-1}{2}$ when $D$ is odd, the character \eqref{eq:charpform} is not the character associated with an irreducible $SO(1,D)$ representation, but a reducible one that is a direct sum of two irreducible chiral representations (classically described by the 1st-order equations $\epsilon\indices{_{\mu_1 \cdots \mu_p}^{\lambda \nu_1 \cdots \nu_p}}\partial_\lambda A_{\nu_1 \cdots \nu_p} = \pm m A_{\mu_1 \cdots \mu_p}$), whose characters are half of \eqref{eq:charpform}.} Unitary of $SO(1,D)$ implies that $m^2>0$, restricting $\Delta$ to fall within the ranges
\begin{align}\label{eq:massUIR}
    \text{Principal series}: & \qquad \Delta = \frac{D-1}{2}+ i \nu \;,  \qquad \nu \in \mathbb{R} \nn\\
    \text{Complementary series}: &\qquad  \Delta = \frac{D-1}{2}+ \nu \;, \qquad  |\nu| < \left|\frac{D-1}{2} -p\right|  \; .
\end{align}
For the scalar case ($p=0$), the Harish-Chandra character has recently been understood in terms of real-time correlators in the static patch \cite{Grewal:2024emf}. A trivial observation is that 
\begin{align}
	\chi^{\Delta,p}(t) = \chi^{\Delta,D-1-p}(t)\; . 
\end{align}

\paragraph{Massless $p$-form}

From \eqref{eq:pformmass}, we see that a massless $p$-form corresponds to $\Delta= D-1-p$ or $\Delta= p$. Some of the degrees of freedom become pure-gauge in the massless limit, which is reflected in the na\"ive character 
\begin{align}\label{eq:naivechar}
	\hat\chi_p(t) = \sum_{k=0}^p (-)^k \binom{D-1}{p-k}   \frac{e^{-\left(p-k\right)t} +e^{-\left(D-1-p+k\right)t}}{\left|1-e^{-t} \right|^{D-1}} = \sum_{k=0}^p (-)^{k+p} \binom{D-1}{k}   \frac{e^{-k t} +e^{-\left(D-1-k\right)t}}{\left|1-e^{-t} \right|^{D-1}} \; . 
\end{align}
Notice that for the $k=0$ term, there is a $O(1)$ term in its expansion in small $e^{-t}$, which according to \cite{Anninos:2020hfj} needs to be subtracted to get the true character \cite{10.3792/pja/1195522333,10.3792/pja/1195523460,10.3792/pja/1195523378}: 
\begin{align}\label{eq:charconvert}
    \chi_p(t) = \hat\chi_p(t) -(-)^p \; . 
\end{align}
This has been shown to agree with the existing results in the literature \cite{10.3792/pja/1195522333,10.3792/pja/1195523460,10.3792/pja/1195523378} for $p=1$ at any $D$ in \cite{Anninos:2020hfj} and  $p=2$ at $D=6$ in \cite{David:2021wrw}. It is interesting to note that the characters satisfy the duality 
\begin{align}
    \chi_p(t) = \chi_{D-2-p}(t) \; . 
\end{align}

\subsubsection{(Quasi)canonical bulk partition function}\label{sec:Zbulk}

A key observation in \cite{Anninos:2020hfj} is that the Fourier transform
\begin{align}\label{introeq:doschar}
    \tilde\rho^\text{dS}(\omega)\equiv \int_{-\infty}^\infty \frac{dt}{2\pi}e^{i\omega t} \chi \left( \frac{t}{R}\right)
\end{align}
can be interpreted as a spectral density for the single-particle static patch Hamiltonian.\footnote{The precise sense in which \eqref{introeq:doschar} is a spectral density was clarified and extended to the cases of static BTZ and Nariai black holes in \cite{Law:2022zdq,Grewal:2022hlo}.} With this one can define a thermal canonical partition function
\begin{align}\label{appeq:idealgas}
    \log \mc{Z}_{\rm bulk} (\beta)\equiv \log \widetilde \Tr \, e^{-\beta R \hat H} \equiv -\int_0^\infty d\omega \, \tilde\rho^\text{dS}(\omega) \log \left(e^{\frac{\beta R\omega}{2} }-e^{-\frac{\beta R \omega}{2} } \right) 
\end{align}
for free bosonic fields in a static patch at any inverse temperature $\beta$. We have normalized $\beta$ such that the de Sitter temperature corresponds to $\beta=2\pi$. Substituting \eqref{introeq:doschar} into \eqref{appeq:idealgas} gives the formula
\begin{align}\label{appeq:Zbulk}
    \log \mc{Z}_{\rm bulk}(\beta) = \int_0^\infty \frac{dt}{2t}\frac{1+e^{-\frac{2\pi t}{\beta R}}}{1-e^{-\frac{2\pi t}{\beta R}}} \chi \left( \frac{t}{R}\right) \; . 
\end{align}
This integral is UV-divergent in the region $t\to 0$, which can be regularized by constructing a zeta function 
\begin{align}
    \zeta(z)= \frac{1}{\Gamma(z)} \int_0^\infty \frac{dt}{2t} \left( t \mu\right)^z \frac{1+e^{-\frac{2\pi t}{\beta R}}}{1-e^{-\frac{2\pi t}{\beta R}}}  \chi \left( \frac{t}{R}\right) \;. 
\end{align}
This takes the same form as \eqref{eq:integralzeta} upon rescaling $t\to t/\mu$. Subsequently, one can extract physically unambiguous parts of \eqref{appeq:Zbulk} as in \eqref{eq:charintphysical}.

\subsection{Laplacians for transverse $p$-forms on $S^{D}$}

The eigenvalues and degeneracies of the Laplacian $\Delta^T_{p}\equiv d d^\dagger+d^\dagger d $ acting on a transverse $p$-form ($1\leq p\leq D-1$) on a round $S^D$ with radius $R$ are \cite{pformspec,Copeland:1984qk,Elizalde:1996nb} 
\begin{align}\label{eq:degeneracy}
	\lambda^D_{n,(p)} &= \frac{\left( n+p\right) \left(n-p+D-1 \right)}{R^2} \; , \quad \nn\\
 d^D_{n,(p)}
 &=\frac{(D-1+2 n) \Gamma (D+n)}{(n+p) (D-1+n-p) \Gamma (n)  \Gamma (p+1)  \Gamma (D-p)} \; ,
\end{align}
where $n\geq 1$. For our discussion in Appendix \ref{app:sphere_PI}, we would like to extend \eqref{eq:degeneracy} to integers below $n=1$, defined as limits. Doing so, we see that
\begin{align}
	\lambda^D_{n,(p)} =\lambda^D_{-D+1-n,(p)} \;, \qquad d^D_{n,(p)} = d^D_{-D+1-n,(p)}
\end{align}
and 
\begin{align}
	d^D_{n,(p)} = 0 \qquad \text{for} \qquad -D+1\leq n \leq 0
\end{align}
except $n=-p$ or $n=-D+1 + p$, where 
\begin{align}\label{appeq:degen-p}
    d^D_{-p,(p)} \equiv (-)^p \; , \quad d^D_{-D+1 + p,(p)} \equiv (-)^{D-1 - p} \; . 
\end{align}
for $p\neq \frac{D-1}{2}$, while
\begin{align}\label{appeq:degenspecial}
    d^{2p+1}_{-p,(p)} \equiv 2(-)^p \; . 
\end{align}
When obtaining the bulk-edge split of $p$-form path integrals, we make heavy use of the identity
\begin{align}\label{eq:degiden}
	d^D_{n,(p)} =\binom{D-1}{p} d^D_{n,(0)} -d^{D-2}_{n+1,(p-1)} \;.
\end{align}
Note that this is true even for $n\leq 1$.

For discussions about dualities, it is useful to note that the eigenvalues and degeneracies \eqref{eq:degeneracy} are invariant upon sending $p\to D-1-p$, i.e.
\begin{align}\label{appeq:eigenduality}
    \lambda^D_{n,(p)} =\lambda^D_{n,(D-1-p)} \;, \qquad d^D_{n,(p)} = d^D_{n,(D-1-p)} \; . 
\end{align}

\subsection{Massive $p$-forms on $S^D$}\label{app:massivepformSD}

The path integral for a $p$-form $A_{\mu_1 \cdots \mu_{p}}$ with mass $m^2R^2=\left(\Delta-p\right)\left(D-1-\Delta-p\right)$ on a round $S^{D}$ with radius $R$ is
\begin{gather}\label{appeq:pformPI}
	\mc{Z}^{p,m^2}_{\rm PI} \left[S^D \right]= \int \mathcal{D}A \, e^{-S[A]} \; , \quad 
	S = \int_{S^{D}} \left( \frac{1}{2(p+1)!} F_{\mu_1 \cdots \mu_{p+1}}F^{\mu_1 \cdots \mu_{p+1}}+ \frac{m^2}{2p!} A_{\mu_1 \cdots \mu_{p}}A^{\mu_1 \cdots \mu_{p}}\right) \; , 
\end{gather}
where
\begin{align}
	F_{\mu_1 \cdots \mu_{p+1}} = (p+1)\nabla_{[\mu_1}A_{\mu_2 \cdots \mu_{p+1}]} \; . 
\end{align}
The path integral measure \eqref{appeq:pformPI} involves all local $p$-forms. Explicitly expanding $A = \sum_\lambda c_\lambda f_\lambda$ in some complete orthonormal basis $f_\lambda$ of $p$-forms, we define the measure to be 
\begin{align}
    \mathcal{D}A = \prod_\lambda \frac{dc_\lambda}{\sqrt{2\pi} \mu} \; , 
\end{align}
where $\mu$ is an arbitrary dimensionful parameter to keep the path integral dimensionless. For the rest of the appendix, we will set $\mu=1$ for notational simplicity and restore it when necessary. The factor $\sqrt{2\pi}$ serves to cancel against the $\sqrt{2\pi}$ factors arising from the Gaussian integrals so that the result only contains the functional determinants of the Euclidean kinetic operators (divided by $\mu$) without any extra factors.

In \cite{Anninos:2020hfj}, it was argued that \eqref{eq:pformPI} can be put to the form
\begin{align}\label{eq:pformPIheat}
	\log \mc{Z}^{p,m^2}_{\rm PI} \left[S^D \right] = &\int_0^\infty \frac{dt}{2t} \left(e^{-\left(\frac{D-1}{2}+i\nu \right) \frac{t}{R}}+e^{-\left(\frac{D-1}{2}-i\nu \right)  \frac{t}{R}}\right)
    \sum_{n\in\mathbb{Z}} \Theta \left(\frac{D-1}{2}+n\right)  
    d^D_{n,(p)} e^{-n\frac{t}{R}} \; .
\end{align}
In this expression,
\begin{align}\label{eq:Thetafn}
	\Theta \left(x\right)=
 	\begin{cases}
		1 \; , \qquad & x>0 \\
 		\frac{1}{2} \; , \qquad & x =0 \\
		0 \; , \qquad & x<0
 	\end{cases}  \; ,
 \end{align}
and $d^D_{n,(p)}$ is the degeneracy \eqref{eq:degeneracy} for the $n$-th transverse $p$-form spherical harmonics on $S^D$. Here we focus on the case of the principal series \eqref{eq:massUIR}; the complementary case is obtained by analytic continuation. To derive \eqref{eq:pformPIheat} from \eqref{eq:pformPI}, one decomposes the $p$-form into a transverse and longitudinal part: $A_{\mu_1 \cdots \mu_{p}} = A^T_{\mu_1 \cdots \mu_{p}} + p \nabla_{[\mu_1}C_{\mu_2 \cdots \mu_{p}]}$. The integration over the transverse part $A^T$ leads to the $n\geq 1$ terms in \eqref{eq:pformPIheat}. On the other hand, $C_{\mu_1 \cdots \mu_{p-1}}$ only enters the action through the mass term in the action \eqref{eq:pformPI}; upon integration, this leads to an infinite product of $m^2$. Absorbing this into a local counterterm in a way consistent with locality leads to the $n=-p$ or $n=p-(D-1)$ term in the range $0\geq n \geq -\lfloor \frac{D-1}{2}\rfloor$ in \eqref{eq:pformPIheat}. This is done explicitly for $p=1$ in \cite{Law:2020cpj}.

To proceed, we use the identity \eqref{eq:degiden} in \eqref{eq:pformPIheat}, resulting in the bulk-edge split
\begin{align}
	\log \mc{Z}^{p,m^2}_{\rm PI} \left[S^D \right] =&\log \mc{Z}^{p,m^2}_{\rm bulk}\left[dS_D\right](\beta =2\pi) + \log \mc{Z}^{p,m^2}_{\rm edge}\left[S^D\right]
\end{align}
where
\begin{align}\label{appeq:massiveZbulk}
	\log \mc{Z}^{p,m^2}_{\rm bulk}\left[dS_D\right] = & \int_0^\infty \frac{dt}{2t} \frac{1+e^{-\frac{t}{R}}}{1-e^{-\frac{t}{R}}}\chi \left( \frac{t}{R}\right) \; , \qquad \chi(t) = \binom{D-1}{p}\frac{e^{-\Delta t} +e^{-\bar\Delta t}}{\left|1-e^{-t} \right|^{D-1}}
\end{align}
is the quasicanonical bulk partition function \eqref{appeq:Zbulk} at the inverse de Sitter temperature $\beta =2\pi$ with a massive $p$-form character \eqref{eq:charpform}, and 
\begin{align}\label{eq:massiveZedge}
	 \log \mc{Z}^{p,m^2}_{\rm edge} \left[S^D\right]= &- \int_0^\infty \frac{dt}{2t} \left(e^{-\left(\frac{D-3}{2}+i\nu \right) \frac{t}{R}}+e^{-\left(\frac{D-3}{2}-i\nu \right)  \frac{t}{R}}\right)\sum_{n\in\mathbb{Z}} \Theta \left(\frac{D-3}{2}+n\right)   d^{D-2}_{n,(p-1)} e^{-n\frac{t}{R}} 
\end{align}
is the edge partition function. Notice that the edge path integral is exactly the negative of \eqref{eq:pformPIheat} but with $D\to D-2$ and $p\to p-1$. Thus we conclude that 
\begin{align}\label{appeq:Zedgemassivepform}
	 \mc{Z}^{p,m^2}_{\rm edge} \left[S^{D} \right] =  \frac{1}{\mc{Z}^{p-1,m^2}_{\rm PI} \left[S^{D-2} \right]}\; . 
\end{align}


\subsection{Massless $p$-form gauge fields on $S^D$}
\label{app:masslesspformSD}

In this section, we provide the details for obtaining the bulk-edge split of the partition function \eqref{eq:masslessPIdet} of a massless $U(1)$ $p$-form gauge field on $S^{D}$. For ease of reference, we restate it here:
\begin{align}\label{appeq:masslessPIdet}
 	\mc{Z}^{p}_{\rm PI} \left[S^D\right]
  = \left(\frac{\mu^{p+1}\sqrt{2\pi R^D{\rm Vol}\left(S^D\right)}}{q}\right)^{(-)^p}   \left( \det\hspace{0mm}'\frac{\Delta_{0}}{\mu^2}\right) ^{-\frac{(-)^{p}}{2}}  \prod_{k=1}^{p}\left( \det\frac{\Delta^T_{k}}{\mu^2}\right) ^{-\frac{(-)^{p-k}}{2}} \;. 
\end{align}
We proceed as in the massive case, putting \eqref{appeq:masslessPIdet} into the form
\begin{align}\label{appeq:SDmasslesspformPI}
	\log \mc{Z}^{p}_{\rm PI} \left[S^D\right]
 =& \, \sum_{k=0}^p (-)^{p-k}\int_0^\infty \frac{dt}{2t} \left(e^{-k\frac{t}{R}}+e^{-\left(D-1-k\right)\frac{t}{R}}\right) \sum_{n=0}^\infty d^D_{n,(k)} e^{-n\frac{t}{R}} \nn\\
 & \quad -  (-)^{p}\int_0^\infty \frac{dt}{2t}  \left(1+e^{-\left(D-1\right)\frac{t}{R}}\right)   + (-)^{p}\log \frac{\mu^{p+1}\sqrt{2\pi R^D{\rm Vol}\left(S^D\right)}}{q}\; .
\end{align}
Here the first line is the contribution from putting together the massive result \eqref{eq:pformPIheat} with appropriate values of $\Delta$ associated with the determinants \eqref{eq:masslessPIdet}. Specifically for $\det \Delta^T_{k}$, $\Delta = k$. The first term in the second line serves to exclude the zero mode from the 0-form determinant. We have set $\mu=1$ in the integrals but kept it explicit in the $U(1)$ volume factor.

Now, applying \eqref{eq:degiden} to each term in the $n$-sum in \eqref{appeq:SDmasslesspformPI}, we obtain a bulk-edge split for \eqref{appeq:SDmasslesspformPI}. The contribution from the first terms in \eqref{eq:degiden} leads to the na\"ive quasicanonical bulk partition function (at $\beta=2\pi$)
\begin{align}
	\log \mc{Z}^{\rm naive}_{\rm bulk}\left[dS_D\right]= & \int_0^\infty \frac{dt}{2t} \frac{1+e^{-\frac{t}{R}}}{1-e^{-\frac{t}{R}}}\hat \chi \left(\frac{t}{R} \right) \; .
\end{align}
Here $\hat \chi(t)$ is the na\"ive character \eqref{eq:naivechar}. The true quasicanonical bulk partition function is related to this by 
\begin{align}\label{eq:Zbulktruenaive}
    \log \mc{Z}^p_{\rm bulk}\left[dS_D\right] = & \int_0^\infty \frac{dt}{2t} \frac{1+e^{-\frac{t}{R}}}{1-e^{-\frac{t}{R}}}\chi \left(\frac{t}{R} \right) = \log \mc{Z}^{\rm naive}_{\rm bulk}\left[dS_D\right]  - (-)^p \int_0^\infty \frac{dt}{2t} \frac{1+e^{-\frac{t}{R}}}{1-e^{-\frac{t}{R}}}
\end{align}
with the last term implementing the conversion \eqref{eq:charconvert} of the na\"ive character into a true character $\chi(t)$. Inspired by the Maxwell case, we are led to group all other terms into an edge partition function, i.e.
\begin{align}\label{appeq:Zedgedef}
	 &\log \mc{Z}^p_{\rm edge}\left[S^D\right] \nn\\
  \equiv & \, -\sum_{k=1}^p (-)^{p-k}\int_0^\infty \frac{dt}{2t} \left(e^{-k\frac{t}{R}}+e^{-\left(D-1-k\right)\frac{t}{R}}\right) \sum_{n=0}^\infty d^{D-2}_{n+1,(k-1)} e^{-n\frac{t}{R}} \nn\\
 & \qquad +(-)^p \int_0^\infty \frac{dt}{2t} \frac{1+e^{-\frac{t}{R}}}{1-e^{-\frac{t}{R}}} -  (-)^{p}\int_0^\infty \frac{dt}{2t}  \left(1+e^{-\left(D-1\right)\frac{t}{R}}\right)   + (-)^{p}\log \frac{\mu^{p+1}\sqrt{2\pi R^D{\rm Vol}\left(S^D\right)}}{q} \nn\\
  =& -  \sum_{k=0}^{p-1} (-)^{p-1-k} \int_0^\infty \frac{dt}{2t} \left(e^{-k\frac{t}{R}}+e^{-\left(D-3-k\right)\frac{t}{R}}\right) \sum_{n=0}^\infty      d^{D-2}_{n,(k)}  e^{-n\frac{t}{R}}\nn\\
  & +  (-)^{p-1}\int_0^\infty \frac{dt}{2t}  \left(1+e^{-\left(D-3\right)\frac{t}{R}}\right)     - (-)^{p-1}\log \frac{\mu^{p}\sqrt{2\pi R^{D-2}{\rm Vol}\left(S^{D-2}\right)}}{q} \; . 
\end{align}
In the first equality, the first line comes from collecting the contributions from the second terms upon applying \eqref{eq:degiden} to the $n$-sum in \eqref{appeq:SDmasslesspformPI}; in the second line, the first term comes from the last term in \eqref{eq:Zbulktruenaive} while the other two terms from the second line of \eqref{appeq:SDmasslesspformPI}. In the second equality, we have shifted $n\to n-1$ and $k\to k+1$ in the sums in the first line, with the first term on the second line canceling against the term with $k=n=0$; we have also evaluated
\begin{align}\label{appeq:evaluate}
    \int_0^\infty \frac{dt}{2t} \left(\frac{1+e^{-\frac{t}{R}}}{1-e^{-\frac{t}{R}}} - 1-e^{-\left(D-1\right)\frac{t}{R}}\right) \Bigg|_{\rm UV-finite}= \frac12\log \frac{D-1}{2\pi \mu R}
\end{align}
with the prescription \eqref{eq:charintphysical} and used the fact that ${\rm Vol}\left(S^{D}\right)=\frac{2\pi}{D-1}{\rm Vol}\left(S^{D-2}\right)$. Now, observe that \eqref{appeq:Zedgedef} is exactly the negative of \eqref{appeq:SDmasslesspformPI} with $D\to D-2$ and $p\to p-1$, which leads us to conclude
\begin{align}\label{appeq:masslessPIbulkedge}
    \mc{Z}^{p}_{\rm PI} \left[S^D\right] = \mc{Z}^p_{\rm bulk}\left[dS_D\right]\mc{Z}^p_{\rm edge}\left[S^D\right] \; , \qquad  
    \mc{Z}^p_{\rm edge}\left[S^D\right] = \frac{1}{\mc{Z}^{p-1,U(1)}_{\rm PI} \left[S^{D-2}\right]} \; . 
\end{align}
As a final note in this appendix, when $p\leq \frac{D}{2}$, this expression can be iterated as
\begin{align}\label{appeq:masslessPIbulkedgeiterate}
\mc{Z}^{p}_{\rm PI} \left[S^D\right] = \frac{\mc{Z}^p_{\rm bulk} \left[ dS_D\right]}{\mc{Z}^{p-1}_{\rm bulk} \left[ dS_{D-2}\right]} \frac{\mc{Z}^{p-2}_{\rm bulk} \left[ dS_{D-4}\right]}{\mc{Z}^{p-3}_{\rm bulk} \left[ dS_{D-6}\right]} \cdots \left( \frac{\mc{Z}^{0}_{\rm PI} \left[S^{D-2p}\right]}{\mc{Z}^1_{\rm bulk} \left[ dS_{D-2p+2}\right]}\right)^{(-)^p} \;,
\end{align}
where $\mc{Z}^i_{\rm bulk} \left[ dS_j\right]$ are defined as in \eqref{eq:Zbulktruenaive}, and 
\begin{align}
	 & \log \mc{Z}^{0}_{\rm PI} \left[S^{D-2p}\right] \nn\\
 =&\, \int_0^\infty\frac{dt}{2t}\frac{1+e^{-\frac{t}{R}}}{1-e^{-\frac{t}{R}}} \left[\frac{1+e^{-(D-2p-1) \frac{t}{R}}}{\left( 1-e^{-\frac{t}{R}}\right)^{D-2p-1}} -1\right] + \log \frac{\sqrt{2\pi R^{D-2p-2}{\rm Vol}\left(S^{D-2p-2}\right)}}{q}
\end{align}
is the partition function for a compact scalar on a round $S^{D-2p}$ of radius $R$ with a target $U(1)$ circle of radius $\frac{2\pi}{q}$ \cite{Law:2020cpj,Ball:2024hqe}. The iterated version \eqref{appeq:masslessPIbulkedgeiterate} (up to the correct zero mode contribution) was presented in \cite{David:2021wrw}.

\bibliographystyle{utphys}
\bibliography{ref}

\end{document}